\documentclass[namedreferences]{solarphysics}
\usepackage[hyperref,optionalrh,solaromanenum]{spr-sola-addons}  
\usepackage{graphicx}                    
\usepackage{color}                       
\usepackage{breakurl}                       

\usepackage{xcolor}                       
\usepackage{siunitx}
\usepackage{gensymb}
\newcommand*{\rom}

\begin{document}

\begin{article}

\begin{opening}
\title{CME Productive and Non-productive Recurring Jets Near an Active Region AR11176}

\author[addressref=aff1,corref,email={ritikas.rs.phy15@itbhu.ac.in}]{\inits{Ritika}\fnm{Ritika}~\lnm{Solanki}\orcid{https://orcid.org/0000-0003-3266-1746}}
\author[addressref=aff1]{\inits{Abhishek}\fnm{Abhishek K.}~\lnm{Srivastava}}
\author[addressref=aff1]{\inits{Bhola}\fnm{Bhola N.}~\lnm{Dwivedi}}
\address[id=aff1]{Department of Physics, Indian Institute of Technology (BHU), Varanasi - 221005, India.}
\runningauthor{R. Solanki \textit{et al.}}
\runningtitle{Recurring Jets Near AR11176}
\begin{abstract}
We study the recurring jets near AR11176 during the period 2011 March 31 17:00 UT to April 1 05:00 UT using observations from the \textit{Atmospheric Imaging Assembly} (AIA) on board \textit{Solar Dynamics Observatory} (SDO). Mini-filaments (mini-filament1 \& 2) are found at the base of these recurring jets where mini-filament1 shows the partial signature of eruption in case of Jet1-3. However, the mini-filament2 shows a complete eruption driving a full blow-out jet (Jet4). The eruption of mini-filament2 triggers C-class flare and Jet4 eruption. The eruption of Jet4 triggers a coronal mass ejection (CME). The plane-of-sky velocity of recurring jets (Jet1-4) results in \SI {160} {\km\per\s}, \SI{106} {\km\per\s}, \SI{151} {\km\per\s} and \SI{369} {\km\per\s}. The estimated velocity of CME is \SI {636} {\km\per\s}. The continuous magnetic flux cancellation is observed at the base of the jet productive region which could be the cause of the eruption of mini-filaments and recurring jets. In the former case, the mini-filament1 shows partial eruption and first three jets (Jet1-3) are produced, but the rate of cancellation was rather low. However, in the latter case, mini-filament2 fully erupts (perhaps because of a higher cancellation rate), and this triggers a C-class flare and a CME-productive jet. At the base of first three jets (Jet1-3), the mini-filament1 causes to push the overlying dynamic complex thin loops resulting in the reconnection and drives the jet (Jet1-3) eruptions. The formation of the plasma blobs are observed during the eruption of the first jet (Jet1). 
\end{abstract}
\end{opening}
\section{Introduction}
Solar transient events are the magnetically driven explosions in the Sun's atmosphere, which carry out the magnetized plasma and inherent energy from the lower to the upper regions of the atmosphere. These eruptions can be classified into two categories, first one is the collimated and confined ejection of a variety of the coronal jets. Second class of the eruptions is known as large scale eruptions like flares and CMEs. The coronal jets are hot and/or cool collimated plasma eruptions along open/curved magnetic fields in the solar atmosphere. The jet phenomenon has been first introduced as X-ray jet by \citet{1992PASJ...44L.173S} and \citet{1992PASJ...44L.161S} in the base-line of the \textit{Soft X-ray Telescope} (SXT)/\textit{Yohkoh} observations. These plasma eruptions occur in quiet Sun (QS), active region (AR) and coronal holes (CH) \citep [\textit{e.g.,}] [and references cited there] {1996PASJ...48..123S,2007PASJ...59S.771S,2009SoPh..259...87N,2010ApJ...720..757M,2013ApJ...769..134M,2016SSRv..201....1R} thereby play an important role in mass and energy transport there. 
The apparent length and breadth of the coronal jet is about $10^{4}$-$10^{5}$ km and $10^{3}$-$10^{4}$ km \citep{1996PASJ...48..123S} as estimated in a variety of EUV imaging observations. The jet's spire moves with an apparent velocity of $10$ to \SI {1000}{\km\per\s} and transports the kinetic energy of $10^{25}$ to $10^{28}$ ergs \citep{1992PASJ...44L.173S} as estimated from the kinematics in 2D-image data. These confined transient phenomena may have importance in energizing a variety of solar coronal regions and transporting the mass to nascent solar wind \citep{2016SSRv..201....1R}. Coronal jet's plasma covers a broad spectrum of emissions and absorption lines based on their wavelengths and are classified as X-ray jets, EUV jets and H$\alpha$ surges exhibiting different temperature regimes \citep [\textit{e.g.,}] [] {2007Sci...318.1591S,2015Natur.523..437S}. The H$\alpha$ surges and coronal jets nearly have the same morphological structure, but their physical properties are completely different. The H$\alpha$ surges (cool jets) are seen as absorption features as observed on the solar disk, which are usually visible in H$\alpha$ and other chromospheric/transition region emissions \citep{2000SoPh..196...79S}. These cool surges are associated with the flare, filament formation and its eruption and coronal mass ejection (CME) in the Sun  \citep[\textit{e.g.,}][and references cited there]{2005ApJ...631L..93L,2005ApJ...628.1056L,2009SoPh..255...79C,2011ApJ...738L..20H}. Solar surges have been modeled by the flux emergence and magnetic reconnection process \citep{1995Natur.375...42Y}, impulsive pressure pulse and explosive events \citep [\textit{e.g.,}] [and references cited there]{1979SoPh...63..187S, 1982SoPh...77..121S,2013ApJ...763...24K,2019AnGeo..37..891S}. Apart from hot/cool coronal jets and cool H$\alpha$ surges, there exist a different class of giant rotating plasma structures which result in coronal swirling motion and tornadoes \citep [\textit{e.g.,}] [and references cited there]{2012Natur.486..505W,2014ApJ...785L...2S,2014PASJ...66S..10W}. These fast-rotating structures channelize the energy from lower to upper layers of the solar atmosphere. The chromospheric swirl was firstly discovered analyzing the  Swedish 1 m Solar Telescope \citep [SST;][]{2003SPIE.4853..341S} observations as a dark ring in Ca {\sc ii} line (854.2 nm). These chromospheric swirls are the result of the rotation of magnetic structure.
The visualization of the flow field of the plasma produces spiral velocity streamlines, which exhibit the features of the magnetic tornadoes.
\par
There are some models which signify the eruption and triggering of coronal jets, \textit{e.g.} the one is known as ``emerging flux model". This model is based on the magnetic reconnection process in a typical magnetic field configuration \citep{1992PASJ...44L.173S}. In this model, as the magnetic bipole emerges in a unipolar magnetic environment, the magnetic reconnection at the null point (X-point) above the bipolar field takes place which triggers the jet plasma motions. Thereafter, the heated plasma is transferred to the open field lines in the outward direction as well as the closed field lines in the downward direction. The hot plasma goes along the open magnetic field lines and forms a coronal jet eruption and open magnetic field of the jet can be a part of the open interplanetary field or large-scale loops \citep{1996ASPC..111...29S,2007PASJ...59S.745S}. The hot plasma is also transferred to the closed field in the downward direction and forms a brightened flare-like loop structure, which is sometimes called as a jet bright point (JBP) at its base. Another mechanism of the triggering of coronal jet is a mini-filament eruption model as proposed by \citet{2015Natur.523..437S}. In this model, the mini-filament eruption is found as a driver of the coronal jets. The jet bright point is created by the reconnection of the closed mini-filament carrying the field. This miniature version of the mini-filament eruption produces jets likewise as the large filament eruption produces coronal mass ejections (CMEs). Sometimes, the large-scale plasma motions in the form of reconnection driven jets also form the coronal mass ejections (CMEs) in the outer solar atmosphere \citep[\textit{e.g.,}][]{2012ApJ...745..164S,2016ApJ...823..129A,2018ApJ...869...39M,2019ApJ...877...61M,2019SoPh..294...68S,2019ApJ...881..132D}. \par
Various physical processes are at work which are responsible for the coronal jet eruption \textit{e.g.,} magnetic reconnection, flux emergence, and generation of the pressure pulse, and chromospheric evaporation \citep[\textit{e.g.,}] [and references cited there]{1982SoPh...77..121S,1996PASJ...48..353Y,2007Sci...318.1591S,2000SoPh..196...79S,2013ApJ...763...24K,2013ApJ...770L...3K,2018Ap&SS.363..233S,2019PASJ...71...14L}. The magnetic flux cancellation is one of the favourable candidates for the eruption of the coronal jets in all the solar environments such as quiet Sun, coronal hole and the active regions. In the observed events of \citet{2016ApJ...822L..23P,2017ApJ...844..131P}, 10 to 15 quiet-region jets are found to be associated with the flux cancellation at their base. \citet{2018ApJ...853..189P} have analyzed the coronal hole jets and found the signature of the magnetic cancellation at the time of the evolution of these jets. \citet{2016ApJ...821..100S,2017ApJ...844...28S} have also studied the active region jets and observed the flux cancellation signature at their base.\par
The plasmoids or blob-like structures are also reported in the spire during the evolution of the coronal jets as seen in observations and models \citep[\textit{e.g.,}][]{2014A&A...567A..11Z,2016SoPh..291..859Z,2017ApJ...834...79Z,2017ApJ...841...27N}. These localized plasma blobs are formed at the junction of the jet-plasma spire and the base-arch field. The plasma blobs are the magnetic islands that are present in the current sheet and evolve in the jet eruption at the time of the magnetic reconnection. They are formed by the tearing mode instability \citep{1963PhFl....6..459F,2014A&A...567A..11Z,2015ApJ...813...86I}. \citet{2016ApJ...827....4W} have found the formation of such blobs due to tearing mode instability in 3D numerical simulation of the jets. The velocity of the plasmoids is found in good agreement with the reconnection rate \citep{2009ApJ...690..748N}. \citet{2012ApJ...759...33S} have also found the plasmoid with a size of ~0.1 Mm in the chromospheric anemone jets. Sometimes the plasma blobs are ejected quasi-periodically within the large-scale coronal streamers \citep{2009SoPh..258..129S}.  \par
These confined jets sometimes trigger large scale eruption of the magnetized plasma \textit{i.e.,} coronal mass ejection (CME). Usually 
the coronal mass ejections (CMEs) are more often associated with the solar flares and driven by the flux-rope and filament-like magnetic structures. Generally, the filament associated CMEs show the typical bubble-like (loop-like) three-part structures as a bright central core, dark cavity and the bright front of the CME. The bright core of the CME is made-up of the cool and dense filamentary material. The velocity of CMEs ranges from \SI {20} {\km\per\s} to \SI {2000} {\km\per\s} \citep{2004ASPC..325..401Y}. CMEs driven by coronal jets have a different magnetic field structuring as coronal jets propagate along the open magnetic field lines \citep{1996ASPC..111...29S} while typical CMEs are connected along the closed helical field lines \citep{1997ApJ...490L.191C}. A large scale jet can be extended to the outer corona in the form of a narrow CME \citep{1998ApJ...508..899W,2008SoPh..249...75L}. According to the twin CME model of \citet{2012ApJ...745..164S}, a single blow-out jet drags two CMEs in the solar atmosphere as jet-like CME and bubble-like CME. This model is supported by many observational events \citep[\textit{e.g.,}][]{2016ApJ...823..129A,2018ApJ...869...39M,2019ApJ...877...61M,2019SoPh..294...68S,2019ApJ...881..132D}. The study, observations, and analysis of these coupled events (Jet and CME) can give better insight to understand the multi-scale eruptive phenomena in the solar atmosphere and their relation in the interplanetary space. \par
In this research paper, we analyze the behavior of recurrent jets found near an active region NOAA AR 11176 (S17W51) as observed on March 31, 2011 and April 01, 2011. \textcolor{black} {Basically this jet productive region is a magnetically enhanced supergranular cell/magnetic network element above which all these recurring jets are observed.} In these jets, first three jets (Jet1-3) are homologous, but the last one (Jet4) is originated from another location. Jet4 is accompanied with GOES C-class flare and a coronal mass ejection (CME). All the jets are found to be associated with mini-filament eruption. We describe the morphological and physical behavior of these observed jets. We also study the difference of the non-CME producive (Jet1-3) and CME productive (Jet4) coronal jets. We also describe the triggering mechanism of these recurrent jets. In Section 2, we describe observational data and analyses. Observational results are elaborated in Section 3. In the last section, we discuss and conclude our observational findings in the context of observed four recurrent jets.
\section{Observational Data and Analyses}
The following data are used for the analysis of the observed recurrent jets and associated CME.\par

i) \textit{Solar Dynamics Observatory} (SDO)/\textit{Atmospheric Imaging Assembly} (AIA) observations \par

The \textit{Atmospheric Imaging Assembly} \citep[AIA;][]{2012SoPh..275...17L} is an instrument on board \textit{Solar Dynamics Observatory} \citep[SDO;][]{2012SoPh..275....3P} which observes the Sun's transition and coronal regions using multi-temperature filters with pixel width $0.6^{\prime\prime}$. It has a spatial and temporal resolution of $1.5^{\prime\prime}$ and 12 sec respectively.  It has 7 EUV filters as 304 \AA~ (He \textrm{II}), 171 \AA~ (Fe \textrm{IX}), 193 \AA~ (Fe \textrm{XII}, \textrm{XXIV}), 211 \AA~ (Fe \textrm{XIV}), 335 \AA~ (Fe \textrm{XVI}), 131 \AA~ (Fe \textrm{VIII}, \textrm{XXI}), 94 \AA~ (Fe \textrm{XVIII}), which covers the temperature range from $6 \times 10^{4} $ K to $2 \times 10^{7}$ K. We download the SDO/AIA data from JSOC data center \footnote{See \url{http://jsoc.stanford.edu} to download the SDO/AIA data.} over the period from 2011 March 31 17:00 UT, to April 1 05:00 UT.

ii) \textit{Solar Dynamics Observatory} (SDO)/\textit{Helioseismic Magnetic Imager} (HMI)
Observations \par

\textit{Helioseismic Magnetic Imager} \citep[HMI;][]{2012SoPh..275..207S} is another instrument on board SDO which observes the photospheric magnetic field in 6173 \AA~ (Fe \textrm{I}) spectral line with $1^{\prime\prime}$ spatial resolution and 45 sec temporal resolution. SDO/HMI data is used for analysing the magnatic polarity regions in and around the jets in AR11176. We download the SDO/HMI line-of-sight magnetograms data from the JSOC data center \footnote{See \url{http://jsoc.stanford.edu} to download the SDO/HMI data.} over the period from 2011 March 31 17:00 UT, to April 1 05:00 UT. We align the SDO/HMI data with SDO/AIA by using the hmi\_prep subroutine of SSWIDL \citep{1998SoPh..182..497F}. \par

iii) \textit{Hinode}/ \textit{X-ray Telescope} (XRT) Observations \par

The \textit{X-ray telescope} is the X-ray imager on board Hinode satellite \citep{2007SoPh..243...63G}. It has $2k\times2k$ CCD and $1^{\prime\prime}$ pixel width and it observes the solar plasma between the temperature range 1 MK - 20 MK. We download the Hinode/XRT data from the data archive of Hinode \footnote{See \url{https://xrt.cfa.harvard.edu} to download the Hinode/XRT data.} for analyse the behavior of C-3.1 flare which is accompanied with Jet4. We use XRT Be\_thick filter data, which is sensitive to 10 MK temperature plasma. \par

iv) \textit{Solar Terrestrial Relation Observatory} (STEREO)/\textit{} (SECCHI) Observations \par

\textit{Sun Earth Connection Coronal and Heliospheric Investigation} \citep[SECCHI;][]{2008SSRv..136...67H} is the five telescope packages on board STEREO which are \textit{Extreme Ultraviolet Imager} (EUVI 1-1.7 $R_\odot$), coronagraphs COR1 (1.5-4 $R_\odot$) \& COR2 (2.5-15 $R_\odot$), heliospheric imagers HI1 (15-84 $R_\odot$) \& HI2 (66-318 $R_\odot$). We use STEREO\_A COR2 data from UKSSDC data centre \footnote{See \url{https://www.ukssdc.ac.uk/solar/stereo/data.html} to download the STEREO\_A and STEREO\_B data.} to study the kinematics of Jet4 associated CME.

v) \textit{Solar and Heliospheric Observatory} (SoHO)/\textit{Large Angle Spectrometer observatory} (LASCO) Observations \par

The \textit{Large Angle Spectroscopic Coronagraph} \citep[LASCO;][]{1995SoPh..162..357B} on board \textit{Solar and Heliospheric Observatory} \citep[SoHO;][]{1995SoPh..162....1D} consists three coronagraphs which are C1 (1.1-3 $R_\odot$), C2 (1.5-6 $R_\odot$) and C3 (3.7-30 $R_\odot$). We take the LASCO C2 data from CDAW data centre \footnote{See \url{https://cdaw.gsfc.nasa.gov/CME_list/data} to download the LASCO C2 data.} for the duration 03:00 UT and 07:00 UT on April 01, 2011 to study the behavior of Jet4 associated CME. 

vi) \textit{Geostationary Operational Environmental Satellite} (GOES) Observations \par

\textit{Geostationary Operational Environmental Satellites} \citep[GOES;][] {1994SoPh..154..275G} is operated by the \textit{National Oceanic and Atmospheric Administration} (NOAA) which carries two X-ray sensors (XRS). These X-ray sensors cover two wavelength regions 1 \AA~ - 8 \AA~ and 0.5 \AA~ - 4 \AA~. We analyze the GOES data over the period from 2011 March 31 17:00 UT, to April 1 05:00 UT in 1 \AA~ - 8 \AA~ wavelength region. We extract the information of GOES flares in this duration from XRT flare catalog \footnote{See \url{https:/xrt.cfa.harvard.edu/flare_catalog} to check the GOES flares.}.\par

vii) \textit{Global Oscillation Network Group} (GONG) H$\alpha$ Observations \par

We use NSO/GONG H$\alpha$ \citep{2011SPD....42.1745H} observations to analyse the behavior of mini-filaments observed at the base of homologous jets (Jet1-3) and Jet4. The NSO/GONG ground-based observatories observe the full disk Sun in 6563 \AA~ H$\alpha$ emissions with 1 minute temporal cadence.
\par

We utilise standard procedures available in the solar-soft \citep{1998SoPh..182..497F} to reduce and calibrate different observational data sets as described in this section.
\section{Observational Results}
\begin{table*}
\caption{Description of the eruption of recurring jets, flare, and, CME in the active region AR11176.}
\label{T-complex}
\begin{tabular}{lcccccc r@{.}l c}
\hline
S.No. & Date & Time Onset (in UT)  & Flare & Jet & CME \\
\hline
i      &  31-03-2011   &       18:49:08       &   --  & Jet1  &  --  \\
ii     &  31-03-2011   &       21:44:08       &   --  & Jet2  &  --  \\
iii    &  01-04-2011   &       00:05:32       &   --  & Jet3  &  --  \\
iv     &  01-04-2011   &       03:48       & C-3.1 &  --   &  --  \\
v      &  01-04-2011   &       03:54:20      &   --  & Jet4  &  --  \\
vi     &  01-04-2011   &       04:17       &   --  &  --   &  CME \\
\hline
\end{tabular}
\end{table*}
\subsection{Source Region of the Recurring Jets}
We analyze the behavior of the recurring jets which were observed over the period from 2011 March 31 17:00 UT, to April 1 05:00 UT using Atmospheric Imaging Assembly (AIA) instrument on board Solar Dynamics Observatory (SDO). The source location of first three jets are same as X-cen=$640^{\prime\prime}$ and Y-cen=$-240^{\prime\prime}$ but Jet4 is asscoiated with X-cen=$620^{\prime\prime}$ and Y-cen=$-230^{\prime\prime}$ (\textit{cf.} Figure 1).\par
The base region of these jets is at the south-west direction of a major sunspot of an active region NOAA AR11176 (S17W51) (\textit{cf.} Figures 2-3 top-panel). The base region of these jets is active and brightened during the above mentioned time-period as it produces the multiple jets. \textcolor{black} {Basically these jets are evolved above a well developed supergranular cell and associated magnetic network.} In these jets, one large scale EUV jet (Jet4) is associated with the coronal mass ejection (CME) and a flare is accompanied at its base (\textit{cf.} Figure 2). The other jets (Jet1-3) are non-CME productive in nature, and were also not associated with any flaring activity (\textit{cf.} Figure 1). The mini-filaments are found at the base of Jet1-3 and Jet4. Mini-filament1 is observed at the base of Jet1-3 and mini-filament2 is observed at the base of Jet4 (\textit{cf.} Figure 4). Mini-filament1 partially erupts and help in triggering of Jet1-3 (\textit{cf.} Figure 4,5,8,10). Mini-filament2 fully erupts, and triggers a C-class flare and a full blow-out jet (Jet4) (\textit{cf.} Figure 4,12). We notice, in these CME producing and non-producing jets that the basic difference is their morphological structure, their development, and their driving mechanism. Also, the mini-filament shows partially eruption in case of first three jets (Jet1-3) and full eruption in case of Jet4. So, we describe the dynamics, evolution, and most likely triggering mechanism of these recurring jets in detail in the forthcoming subsections.\par
\subsection{EUV Brightening and Enhanced GOES Flux During Onset of Jet4}
The behavior of C-class flare at the base of the Jet4 is shown in AIA 1600 \AA~ (\textit{cf.} top-left panel of Figure 2) and soft X-ray image (\textit{cf.} top-right panel of Figure 2).
We analyze the pattern of the EUV brightening with respect to the time at the base of the recurring jets over the period from 2011 March 31 17:00 UT, to April 1 05:00 UT in different SDO/AIA filters (\textit{cf.} bottom panel of Figure 2). The selected field of view (FOV) is $30^{\prime\prime} \times 48^{\prime\prime}$ for extracting the EUV intensities as shown by the overplotted box on SDO/AIA 1600 \AA~ image (\textit{cf.} top-left panel of Figure 2). EUV light curves show quiet phase at the eruption of three jets (Jet1, 2, 3). At the time of the fourth EUV jet (Jet4), these light curves show the enhanced peaks. We also study the GOES flux for the whole evolution period of all recurring jets \textit{(cf.} bottom panel of Figure 2). At the base of the Jet4, the GOES C-class brightening is observed at 03:48 UT. GOES light curve shows the X-ray flux in the wavelength range of 1 \AA~ - 8 \AA~. In the GOES flux, there are four peaks which indicate the B-7.0, C-1.6, C-3.4 and C-3.1 flares while the Jet4 is only accompanied with C-3.1 flare (last peak in GOES lightcurve Table 1).  The GOES flares B-7.0, C-1.6 and C-3.4 are associated from the different locations and they are not associated with any observed recurring jet. We have plotted the dashed lines on the both lightcurves at the eruption time of all recurring jets. If we compare the EUV light curves and GOES light curve for the eruption time of the flare C-3.1 and Jet4, these two fluxes show nearly the same behavior in form of enhanced peaks as observed. The EUV fluxes also enhance with certain obvious time delay. This indicate that the last jet is accompanied with a compact C-class solar flare. The full eruption of mini-filament2 leads both the flare brightening and eruption of Jet4.
\subsection{Evolution of Magnetic Field at The Base of Recurring Jets}
We analyze the behavior of the magnetic field to find out the possible causes of these observed recurring jets (\textit{cf.} Figure 3). So for this, we use SDO/HMI line of sight (LOS) magnetograms data. The time-sequential images of the SDO/HMI magnetograms show the magnetic field behavior in and around the triggering site of all recurring jets (\textit{cf.} upper panel of Figure 3). These images of SDO/HMI magnetograms is aligned by using drot\_map program of IDL \citep{1998SoPh..182..497F} with taking the reference file of 17:00 UT. The first three jets (Jet1, Jet2, Jet3) and fourth jet (Jet4) occur from two different neutral lines N1 and N2 respectively. We have plotted the neutral lines N1 and N2 on HMI magnetogram at 18:49:17 UT (see image (a) in Figure 3). The locations of the neutral lines N1 and N2 are respectively ($640^{\prime\prime}$,$-240^{\prime\prime}$) and ($620^{\prime\prime}$, $-230^{\prime\prime}$). The mini-filaments1 \& 2 reside on these neutral lines and take the part in the eruption of jets. \par
The base region of these jets looks like \textcolor{black} {brightened supergranular cell and magnetically enhanced network lying above it.} The base region of these jets is near to the sunspot of the AR11176 in its south-west direction in the quiet Sun. In this base region, the positive polarity plays a major role and is distributed significantly over the network boundary and also in its middle. Negative polarity has a fewer localized features emerged within the supergranular cell/magnetic network. The positive and negative magnetic fluxes are extracted at the base region of these jets for the whole eruption period (\textit{cf.} lower panel of Figure 3). We infer the base region as the wider area at the photosphere where positive polarity and multiple localized negative polarities are  present. We estimate the temporal behavior of the positive and negative fluxes in this area (see box in upper-left panel of Figure 3). \textcolor{black} {The box covers the part of the supergranular cell and the location on it from where the jets are produced.} We have estimated the temporal variation of positive and negative fluxes for whole base region of jets (Jet1-4). The box region consists both the neutral lines where these recurring jets are produced. Figure 3 clearly indicates that the positive polarities at the boundary of supergranular cell form outer ring, while the emerged negative polarity forms the inner ring. The two neutral lines are almost arc-shape, lie between these two opposite polarities.\par
Magnetic flux estimation is based on our understanding that over all flux cancellation is basically subjected due to many localized emergence of negative fluxes, and their cancellation with positive fluxes. These responses are clearly evident in the EUV emissions (\textit{cf.} Figure 4 and 5)  inferring the localized plasma dynamics that trigger the mini-filaments and recurring jets. Magnetic fluxes are extracted from the region of the box which is overplotted on HMI magnetogram (\textit{cf.} Image (a) of the upper panel of Figure 3). We have plotted dashed vertical lines in Figure 3 (bottom panel) at the eruption time of each jet. In Figure 3, we clearly see the behavior of the magnetic cancellation up to the eruption time of the first three jets (Jet1-3) as positive flux shows decline trend and negative flux shows increasing trend. This behavior suggests that emerging negative fluxes cancel with the existing positive flux and this somewhat played the role in triggering the mini-filament1 and the Jet 1, 2 \& 3.
After the eruption of the Jet3, positive flux again starts increasing.  Negative flux also shows now a decreasing behavior. The already emerged small-scale negative fluxes within the supergranular cell/magnetic network is now collectively decreased due to the cancellation with the increasing positive flux. The rate of change of positive flux during the eruption of mini-filament1 and first three jets is $7.78 \times 10^{22}$ $\textrm{Mx}$ $\textrm{hr}^{-1}$ while the rate of change of positive flux during the eruption of mini-filament2 and initiation of C-3.1 flare is $9.59 \times 10^{22}$ $\textrm{Mx}$ $\textrm{hr}^{-1}$. The rate of change of negative flux during the eruption of mini-filament1 and subsequently the triggering of first three jets is $2.17 \times 10^{22}$ $\textrm{Mx}$ $\textrm{hr}^{-1}$. The rate of change of negative flux during the eruption of mini-filament2 and initiation of C-3.1 flare is $3.23 \times 10^{22}$ $\textrm{Mx}$ $\textrm{hr}^{-1}$. We have noticed that the rate of change of positive and negative fluxes during the initiation of C-3.1 class flare is higher compared to the rate of change of positive and negative fluxes during the eruption of first three jets (Jet 1-3). It should be noted that the full eruption of mini-filament2 causes the triggering of flare and subsequently Jet4.
\subsection{Non-CME Producing Jets (Jet1, Jet2, Jet3)}
We identify four recurring jets in the time-sequence images of AIA 304 \AA~ (\textit{cf.} Figure 1)  over the period from 2011 March 31 17:00 UT, to April 1 05:00 UT. In these recurring jets, first three jets (Jet1, Jet2, Jet3) are non-CME productive jets and Jet4 is a CME productive jet. Figure 2 also depicts that Jet4 is accompanied with a C-class solar flare.
\subsubsection{Eruption of Jet1} 
The eruption of the first jet (Jet1) is observed at 18:49:08 UT on 2011, March 31. We examine the behavior of the base region of Jet1 in multi-temperature filters (AIA 1600 \AA~, AIA 304 \AA~, AIA 171 \AA~, AIA 94 \AA~) (\textit{cf.} Figure 5). At the prior time of the eruption of the first jet (Jet 1) two brightened loop-like structures are noticed at the base of the jet (\textit{cf.} Figure 5). Due to the interaction of these brightened loop-like structures, another brightened loop is formed at which the initiation of Jet1 is observed and jet's spire develops above it. Jet1 is evolved in successive stages. First a faint small spire develops at the southern direction of its base, and after this some plasma erupts from northern direction which completely draws the whole plasma and make a full jet spire showing a partial rotational signature (\textit{cf.} Figure 5 and 7). A mini-filament is observed at the base of the jet which shows the partial/confined eruption as some part of this mini-filament erupts out with the jet's spire (\textit{cf.} Figure 4 and 5). The lifetime of the first jet (Jet 1) is about 15 min and it attains 30 Mm height in the initial 5 minutes time duration. The full evolution of Jet1 can be seen in animation Movie\_1. The kinematics of the Jet1 is studied by using the height-time analysis method. We have calculated plane-of-sky velocity of all four recurring jets by using AIA 304\AA~ (He \textrm{II} formation temperature log(T) $\approx{4.7}$) filter of SDO/AIA. In the lower panel of Figure 6, the left panel shows the selected slit path along the spire of the Jet1 and right panel shows the height-time measurement of this jet. The overplotted path is used to evaluate the velocity of the jet. The calculated plane-of-sky velocity of Jet1 is  \SI {160} {\km\per\s} (\textit{cf.} Figure 6).\par
When we analyze the base of the jet forming region in various AIA filters (\textit{cf.} Figure 5), it is evident that fountain-like thin loop-like structures are generated, which are connected at the boundary of the magnetic network/supergranular cell. This magnetic configuration is very complex and highly dynamic/unstable. It is created due to continuous emergence of the small-scale localized negative polarities within the supergranular cell, while its boundary is associated with the positive magnetic polarities (\textit{cf.} top-panel of Figure 3). The flux cancellation at the base as well as interaction of overlying closed field lines have triggered the plasma blobs episodically along the spire of Jet1.
The formation of the plasma blobs are noticed in the eruption of the Jet1 (\textit{cf.} Figure 7).
Figure 7 shows the zoomed view of the eruption of the Jet1. The multiple plasma blobs are continuously evolving in the eruption of the jet and flowing along with the spire of the Jet1. These recurring plasma blobs are formed due to magnetic reconnection in the localized current sheet at the top of the complex system of thin multiple loops. The mini-filament eruption from below this magnetic system tries to push the overlying complex thin loops which go into reconnection. The continuous brightening is observed at the footpoint of the Jet1 which indicates reconnection driven energization of jet's plasma. Thus, the magnetic reconnection drives the Jet1 above the complex loop system evolved in the lower part of the jet productive region.
\subsubsection{Eruption of Jet2}
The eruption of second jet (Jet2) from the same base region as of first jet is observed on March 31, 2011 at 21:44:08 UT. The zoomed view of the base of Jet2 in multi-temperature filters is shown in Figure 8. The multiple brightenings are observed at 21:43 UT at the base of the jet. The plasma spire of the Jet2 follows the same path as the erupting direction of the first jet. Due to the line-of-sight effect, the spire of the jet looks like a faint structure. In the eruption of Jet2, the mini-filament shows the signature of partial eruption as in case of Jet1 (\textit{cf.} Figure 4 and 8). The life-time of this jet is about 10 min and it attains 20 Mm height at the initial time. The calculated plane-of-sky velocity of Jet2 is  \SI {106} {\km\per\s} (\textit{cf.} Figure 9). The evolution of Jet2 can be seen in animation Movie\_2.\par
The formation mechanism of this second jet is also same like the first jet. The evolved fountain like complex loop system surging out from the centre of the supergranular cell due to the emergence of small-scale negative fluxes, connect with the positive polarities at its boundary. Since the flux emergence and cancellation is continuously ongoing process at the photosphere, the mini-filament subsequently now tries to erupt. The overlying thin loop threads become highly dynamic and unstable. Their interaction and reconnection leads the propulsion of Jet2. \par
\subsubsection{Eruption of Jet3}
The initiation time of the third jet (Jet3) is 00:05:32 UT on 01-04-2011 (\textit{cf.} Figure 10). There is still some activity at the base of the third jet (Jet3) before its eruption, as some brightened and dynamic loop-like structures are visible (\textit{cf.} Figure 10). The triggering mechanism of this jet is very similar to the first two jets, which we have explained in subsection 3.4.1 and 3.4.2. The dynamics of third jet (Jet3) is most likely same as of the first two jets. The mini-filament at the base of the Jet3 shows the partial/confined eruption as some part of it leaks out with the jet eruption as in case of first two jets (\textit{cf.} Figure 4 and 10). It initiates at the southern part, and thereafter some plasma is erupted from the northern direction to make the whole spire of the jet. The lifetime of third jet is about 10 min. The calculated plane-of-sky velocity of Jet3 is  \SI {151} {\km\per\s} (\textit{cf.} Figure 11). The complete evolution of Jet3 can be seen in animation Movie\_3.
\par 
\subsection{CME Producing Jet (Jet4)}
The eruption of Jet4 is observed at 03:54:44 UT on April 01, 2011 (\textit{cf.} Figure 12) and this is accompanied with C-class enhancement and a CME. At the initial time of the eruption of the Jet4, its base looks very complex and brightened and there is small magnetic loops all around it (\textit{cf.} Figure 12). The evolution of this jet in multi-temperature filters of SDO/AIA is shown in Figure 11. A mini-filament is found at the base of Jet4. This mini-filament erupts completely and results in the C-class flare eruption and drives a full blow-out jet (\textit{cf.} Figure 4 and 12). The C-class flare results from the reconnection occuring along with the eruption of the field carrying by the minifilament.
A C-class flare C-3.1 initiates at 03:48 UT and peaks at 03:53:42 UT with bulk energy release, and motion of the plasma forming the jet spire is also evident. The plasma spire of the jet develops above the flaring region. The base of C-3.1 flare and Jet4 is same.
The heated plasma material gains kinetic energy and goes higher in the south-west direction. The material of the jet's spire consists of multi-temperature plasma. The velocity of this jet is  \SI {369} {\km\per\s} (\textit{cf.} Figure 13). This high-speed jet attains the height of about 250 Mm at its initial 10 min time duration. The spire of the Jet4 looks like the spray-like structure. The Jet4 attains its maximum height at about 04:20 UT. The spire of the Jet4 disappears, and a very small amount of the plasma of the jet's spire falls down to its base. The full eruption of Jet4 can be seen in Movie\_4.\par
This EUV jet (Jet4) is associated with a CME eruption. The evolution of CME with time is shown in sequential images of the LASCO C2 coronagraph. The initiation time of the CME is 04:17 UT as observed by the LASCO C2 coronagraph when its leading bright front tends to pop-up in its field-of-view. However, its dense core is visible around 04:28 UT in the coronagraphic field-of-view. The spatial and temporal relationship between Jet4 and CME is well correlated.  The direction of the CME and Jet4 are also same. The time difference between these two events is about 33 min 16 s (Jet4-03:54:44UT as seen in AIA 304 \AA~, and core of CME-04:28 UT in LASCO C2 FOV). Since Jet4 extends very high into the corona (250 Mm projected height around 04:20 UT) so it appears quickly as the core of the CME in LASCO C2 Coronagraph at around 04:28 UT (\textit{cf.} Figure 14). However, the bright shock front of the CME evolves a bit earlier around 04:17 UT. This is the response of jet-like eruption in the highly rarefied outer corona. Large scale EUV jet (Jet4) triggers the CME and become its core (wedge-shaped) (\textit{cf.} image (b) of Figure 14). The bulk plasma which is erupted in CME consists of the jet material. The eruption of CME can be seen in Movie\_5.\par
The kinematics of Jet4 associated CME is studied by using the height-time measurement technique. For analyzing the real height of the CME at different times, we use the tie-pointing method of the \citet{2006astro.ph.12649I}. The tip of the CME is tracked simultaneously in two observational data STEREO\_A and LASCO C2 with a triangulation technique (\textit{cf.} top panel of Figure 15). We have tracked the tip of the CME at different times at 04:38:53 UT, 04:49:26 UT, 05:24:05 UT, 05:36:07 UT, 05:48:05 UT, 06:12:11 UT, 06:24:05 UT and 06:36:05 UT. We repeat the measurement of the real height of CME ten times for a particular time to estimate the uncertainty in the measurement of the height of the CME. The calculated velocity of Jet4 associated CME is  \SI {636} {\km\per\s} (\textit{cf.} bottom panel of Figure 15).
\section{Discussion and Conclusions} 
In this paper, we analyze the observations of the recurring jets observed in the quiet-Sun near the major sunspot of AR11176 by analyzing the observational data of SDO/AIA, SDO/HMI, STEREO\_A/COR2 and SoHO/LASCO C2 instruments. We present below some concluding remarks of this observational finding, and also discuss their implications.\par
i) The jets are associated with a quiet-Sun supergranular cell/magnetic network. The negative magnetic field polarities are emerged in the middle of the supergranular cell with time at the base of all these recurring jets. The magnetic cancellation may be the most likely cause for the mini-filament eruptions and the eruption of these jets.\par
ii) The emerged tiny polarities form a very complex low-lying dynamical loop systems. These loop threads interact and reconnect with each other during the eruption of mini-filaments. Due to this process, the reconnection occurs, which leads localized brightening and plasma eruptions in form of the jets.\par 
iii) In the recurring jets, the first jet (Jet1) shows the plasma blob formation during its eruption. These plasma blobs are formed due to the localized reconnection processes between the loop threads. Moreover, some cool plasma blobs are also seen in 1600 \AA~ moving from the lower atmosphere. This indicates the cancellation within the supergranular cell and localized energy release near the photosphere, which drives the plasma along the field lines. In the overlying atmosphere the mini-filament1 eruption, and, the interaction and reconnection of dynamic loop threads also lead overall plasma dynamics, and trigger multiple localized jets above the supergranular cell/magnetic network. Other jets (2 \& 3) are also formed due the similar physical processes.\par
iv) The evolution of Jet4 indicates its association with C-class flare at its base. It also triggers the coronal mass ejection (CME). All these processes are triggered by mini-filament2 eruption.\par
v) The emerged negative flux within the supergranular cell at the base further canceled out by the emerging positive flux. This is due to the fact that the supergranular cell/magnetic network tends to self-organize its original state of relaxed magnetic topology. This further opens up the entire magnetic configuration, and simplify the evolved complex loop-like structures formed initially  due to the emergence of the tiny negative polarities. The magnetic complexity of the system is self-reorganized. Mini-filament2 also erupts and causes the energy release of C-class flare. It also drives a bulky and long EUV jet (Jet4).\par
vi) The Jet4 is triggered due to the eruption of the mini-filament. This mini-filament eruption drives a C-class flare.
The bulk heated plasma of Jet4's spire goes higher and forms a spray-like structure. This spray-like structure of Jet4 extends very quickly and high into the corona and appears as CME core. \par
In the studied recurring jets, the dynamics of the first three jets are quite similar as at the base of these jets there is a complex loop system evolved. The base region of these jets is magnetically very dynamic and complex and the reason behind this is the continuous flux emergence and flux cancellation and thus the eruptions of mini-filaments and subsequent reconnection in overlying loop system.
CME-non-productive and CME-productive jets might depend on the flux-cancellation rate, but more events must be studied before reaching a conclusion about this. In the present base-line, we can argue that the magnetic flux cancellation may helps driving mini-filament eruptions that later drive the various physical processes resulting into the formation of jets.
Jet productions also depend upon the dynamics and internal reconnection of low-lying complex loop systems. Therefore, the present observations suggest the unique way of magnetic flux emergence, their cancellation, and overying interacting magnetic field configurations, which collectively determine the evolution of jets and their capability in CME-producion. Moreover, this observation also explains the capability of coronal jet in CME formation, which is basically evolved above quiet-Sun in the outer corona.\par
In the observational result of \citet{2011ApJ...738L..20H}, a quiet-Sun jet is associated with the eruption of a mini-CME. However, in this observed event the quiet-Sun jet is driven by a mini-filament and the responsible cause for these multiple transients is magnetic flux convergence and cancellation at the photosphere. \citet{2012ApJ...745..164S} have also studied the quiet-Sun coronal jet which is associated with twin CMEs. At the base of this quiet-Sun jet, there was a mini-filament present that lead its eruptions. In these twin CMEs, one jet-like CME is associated with the hot component of the jet, and other bubble-like CME is associated with the cool component (filamentary material) and the triggering agent of this eruption is magnetic cancellation and magnetic reconnection. \citet{2019SoPh..294...68S} have analyzed the observational results of a quiet-Sun blow-out jet accompanied with a circular mini-filament at its base, associated with the twin CMEs eruption. In this study, a jet-like CME was associated with the hot component of jet, and bubble-like CME was associated with the eruption of a mini-filament. The driving factor for this blow-out jet and twin CMEs was magnetic cancellation. \citet{2019ApJ...881..132D} have also observed the twin CMEs (jet-like and bubble-like) generation with the mini-filament driven coronal jets.
In our observational analysis, we find the recurring jets eruption above a supergranular cell/magnetic network.
In these recurring jets, the first three jets shows the partial/confined eruption of the mini-filament with the jet's spire but in the eruption of Jet4 the mini-filament erupts fully and makes full blow-out jet. At the time of the Jet4 the mini-filament eruption drives a C-class flare, this C-class flare is occuring due to reconnection resulting from the field of the mini-filament moving upward. The first three jets are not full blow-out jets while the fourth jet is the full blow-out jet which further forms a coronal mass ejection (CME). The jet itself constitutes the core of the CME, and its interaction into the rarefied outer corona forms the frontal part of the CME. 
\begin{figure*}
\centerline{\includegraphics[width=14cm,angle=0]{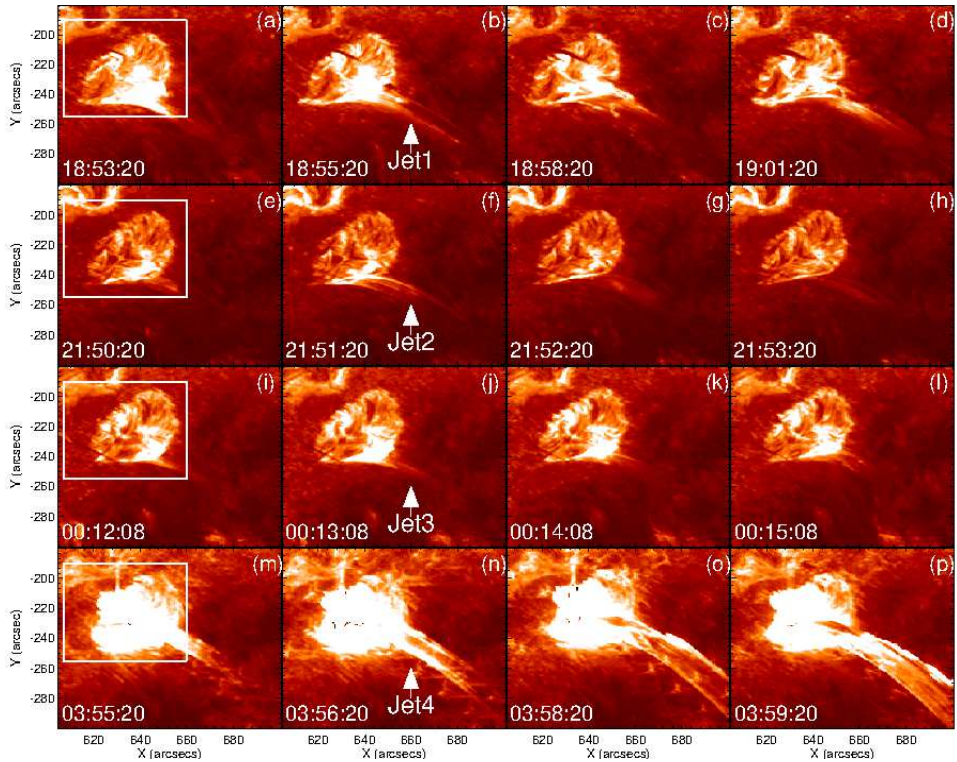}}
\caption{The time-sequential images of the recurring jets in AIA 304 \AA~. This mosaic shows the behavior of Jet1 (a-d), Jet2 (e-h), Jet3 (i-l) and Jet4 (m-p). The initial three jets show nearly same behavior as small loop like structures at their base seem to interact and reconnect. Brightening occurs at the base during the development of their spire. The last jet (Jet4) has the spray like sturcture as it goes faster and attains a maximum height in a short duration time. It is accompanied with a C-class flare, and also yields a CME. At the base of the Jet(1-3) and Jet4 mini-filaments are present which contribute in the eruption of these jets. In Jet1-3 eruption, mini-filament1 partially erupts but in case of Jet4 the mini-filament fully erupts and form a full blow-out jet.}
\label{Figure 1}
\end{figure*}
\begin{figure*}
\centerline{\includegraphics[width=5.3cm,angle=0]{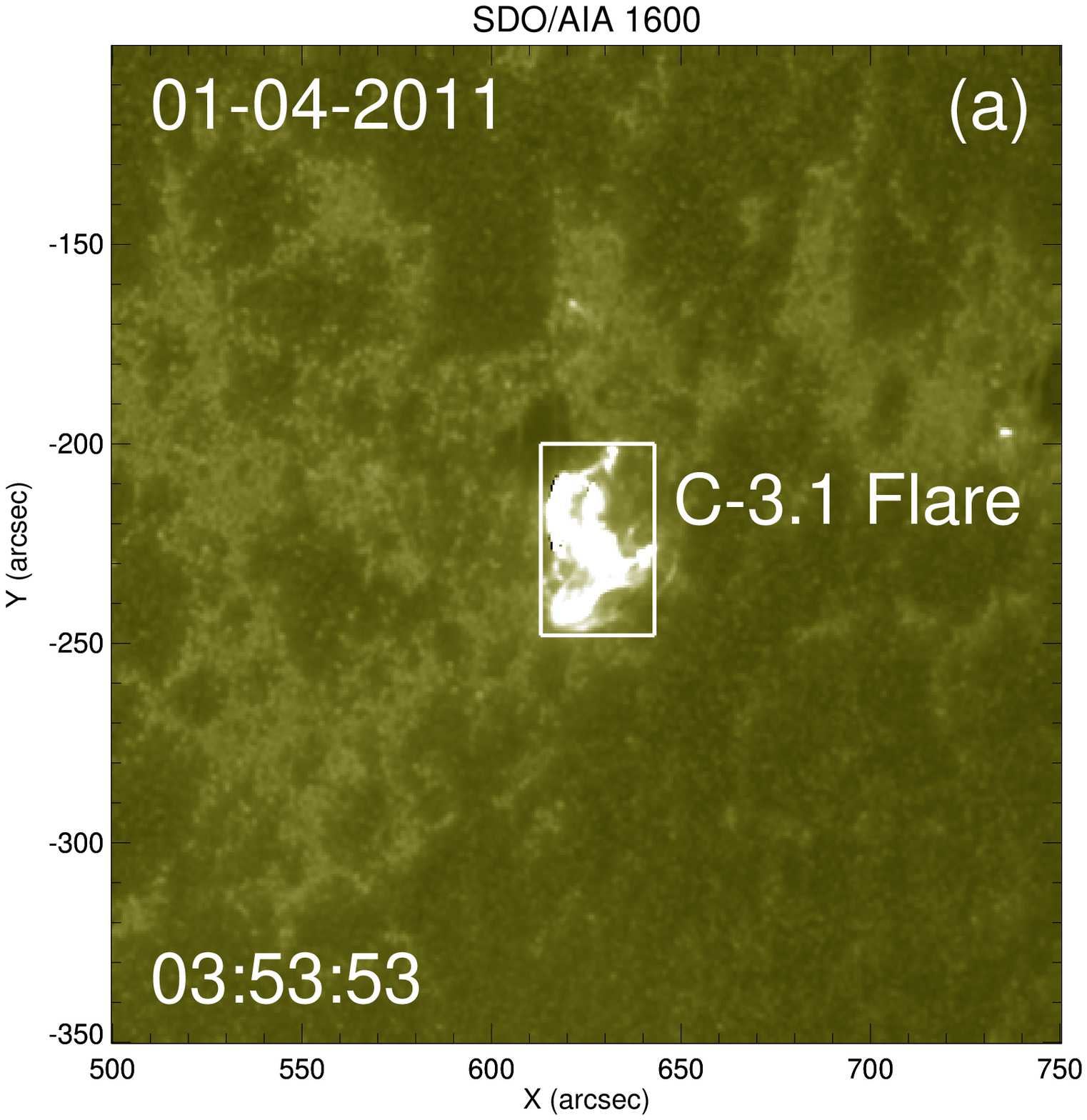}
\vspace*{-0.01\textwidth}
\includegraphics[width=5.3cm,angle=0]{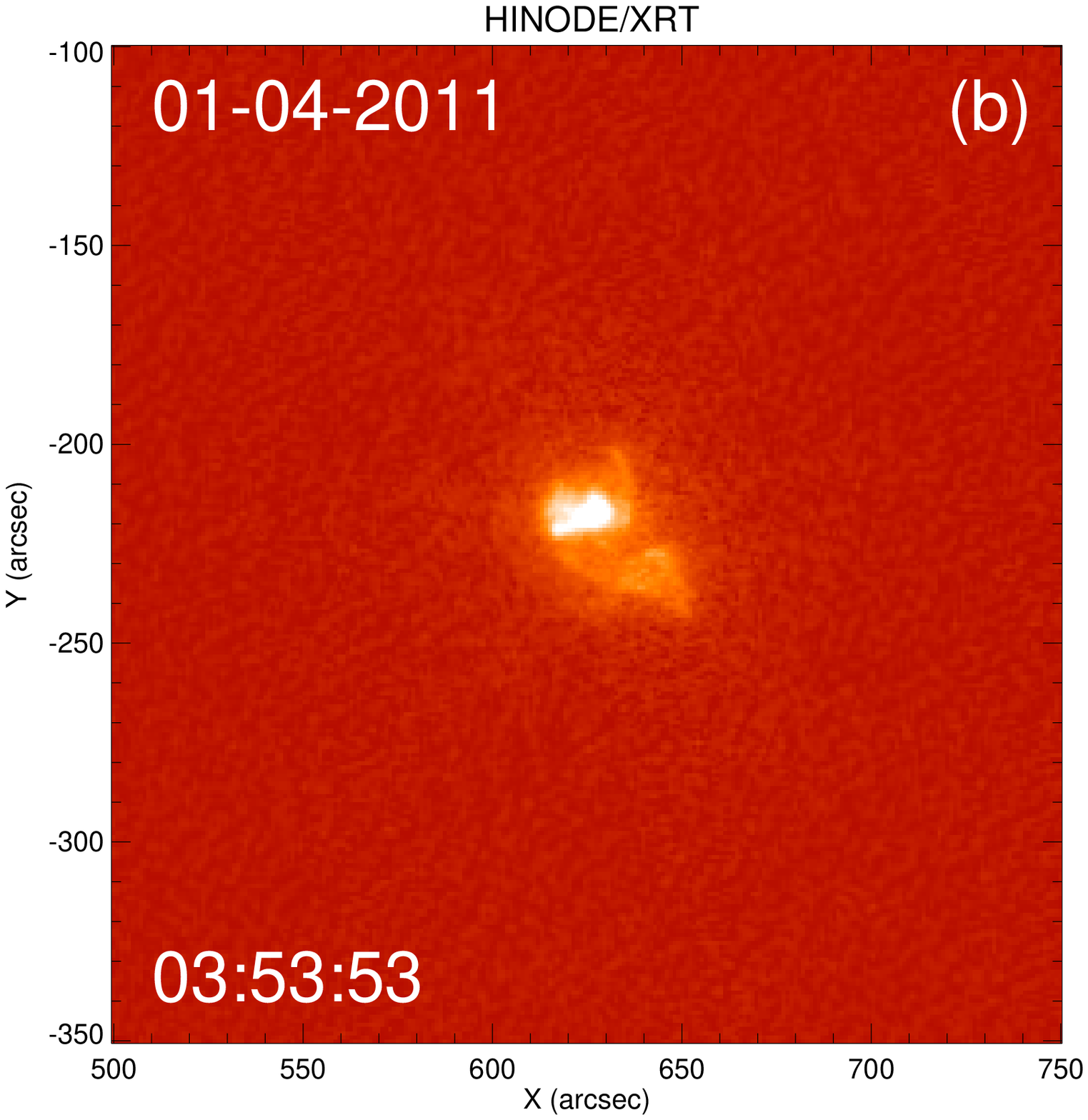}
}
\hspace*{-0.03\textwidth}
\centerline{\includegraphics[width=8.6cm,angle=90]{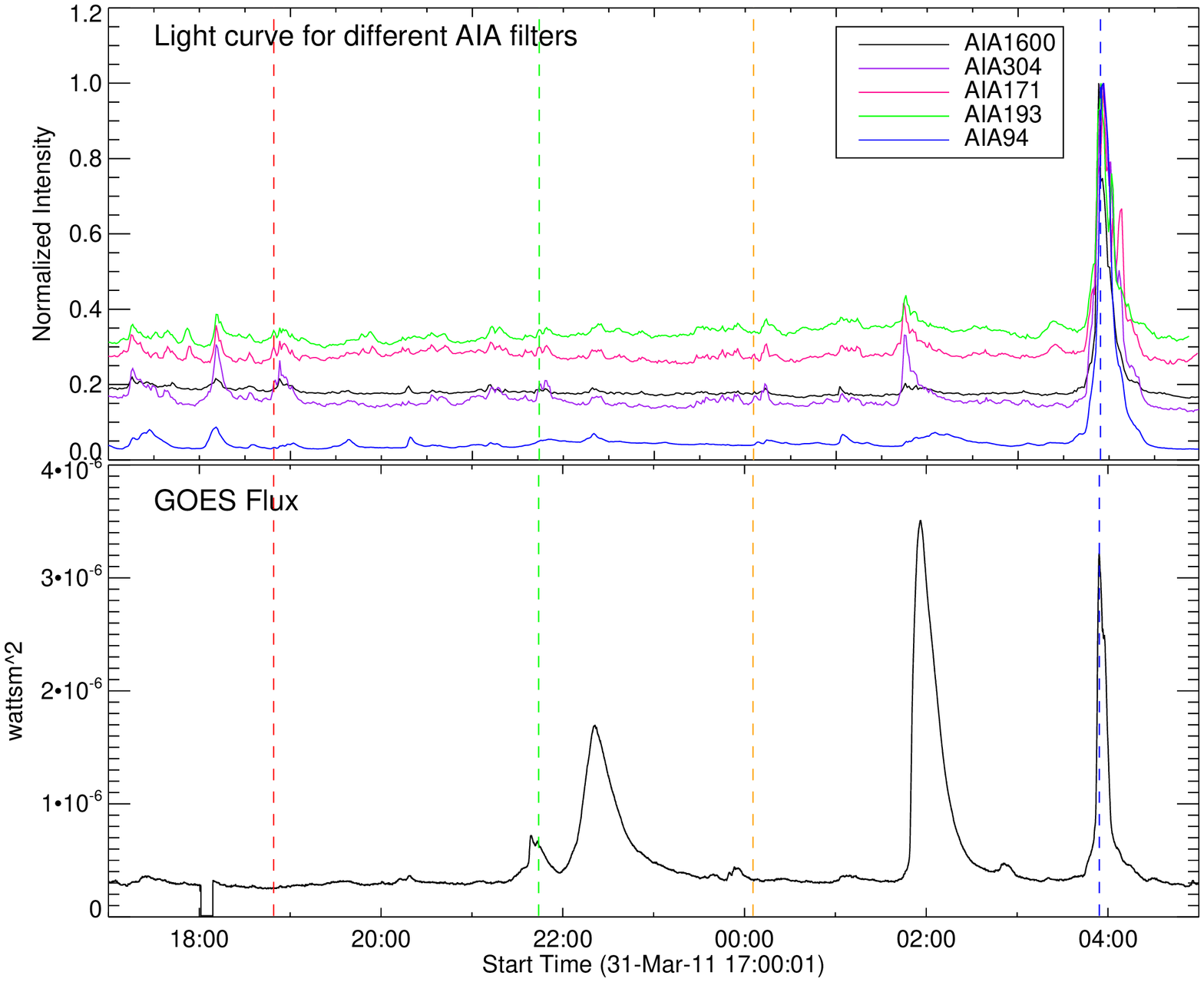}
}

\caption{(a) A C-class enhancement is observed at 03:53:53 UT (peak time of flare) which is shown in AIA 1600 \AA~ image (top left panel) and HINODE/XRT soft X-ray image (top right panel). The mini-filament eruption drives this C-class flare and Jet4 eruption. (b) GOES flux for the wavelength region of 1 \AA~ - 8 \AA~ and the pattern of the EUV brightening at the base of the recurring jets for the whole evolution period in the multiple filters (at multi-temperatures) of SDO/AIA (bottom panel). The EUV intensity is extracted as according to the overplotted box on AIA 1600 \AA~ image (top left panel). The vertical lines on the lightcurve of EUV brightening and GOES flux show the timing of the various jets. The behavior of EUV light curve and GOES light curve for the eruption of the C-3.1 flare and Jet4 matches well to each other.}
\label{Figure 2}
\end{figure*}
\begin{figure*}
\centerline{\includegraphics[width=12.3cm,angle=0,trim={2cm 7cm 2cm 1cm}]{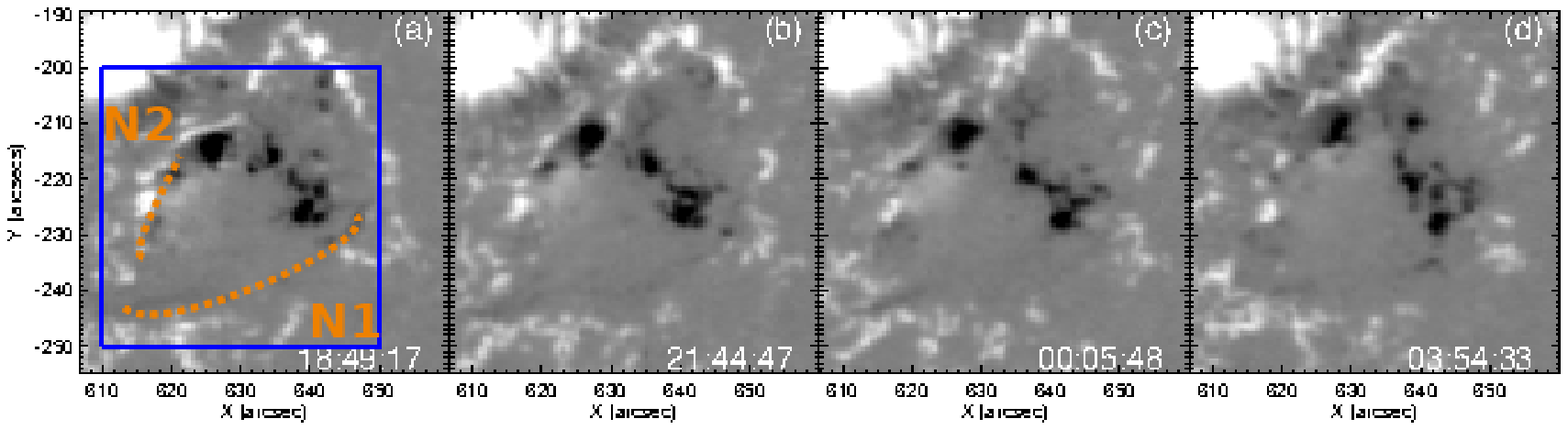}}
\vspace*{-0.01\textwidth}
\centerline{\includegraphics[width=10.5cm,angle=90]{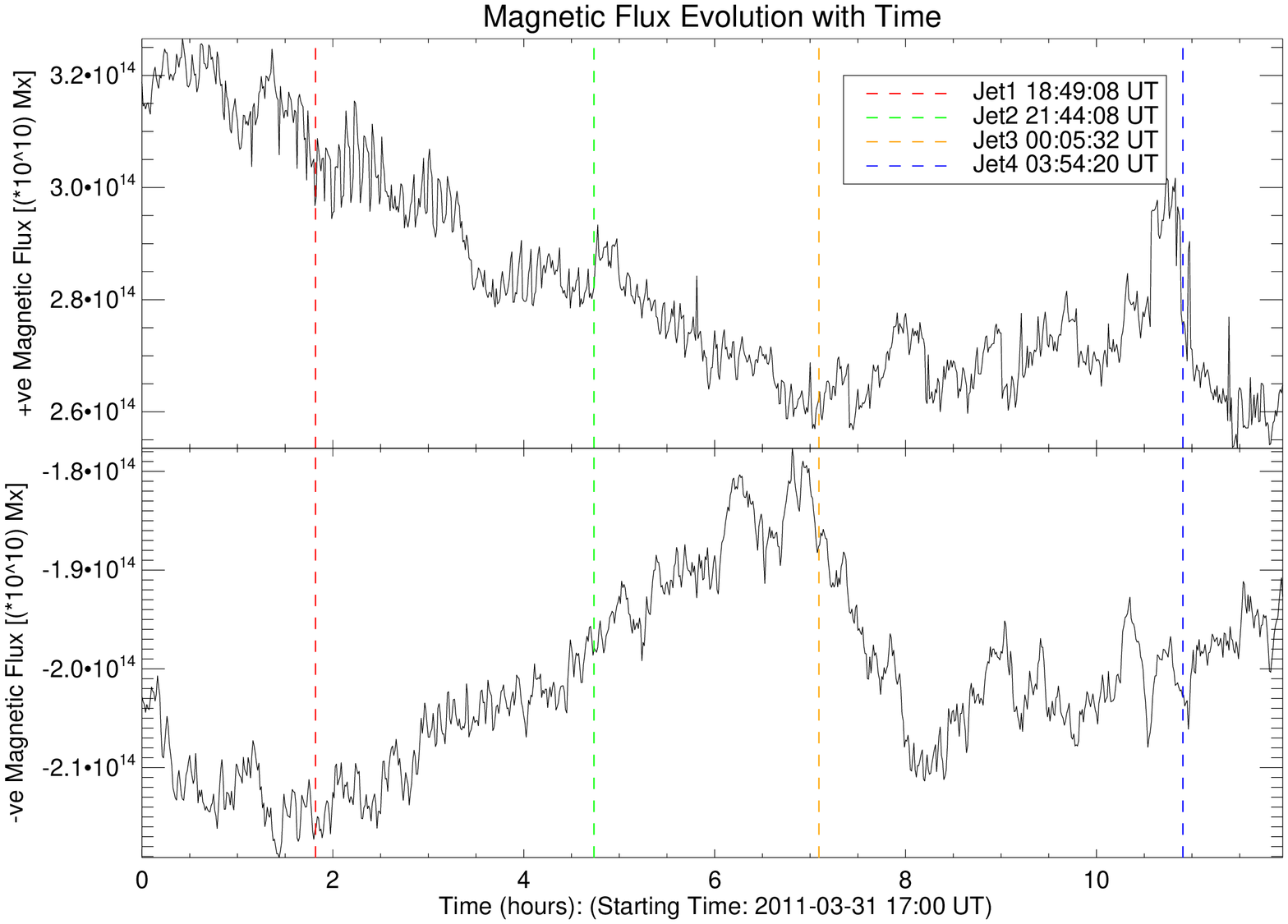}
}
\caption{This figure shows HMI magnetograms to understand the distribution of positive and negative polarities over a supergranular cell/magnetic network from where all recurring jets are triggered. The first three jets (Jet1-3) and fourth jet (Jet4) occur on two different neutral lines N1 ($640^{\prime\prime}$,$-240^{\prime\prime}$) and N2 ($620^{\prime\prime}$, $-230^{\prime\prime}$). N1 and N2 are neutral lines where the mini-filaments reside and contribute in the jet eruption. Mini-filament1 which lies on N1 partially leaks to form first three jets (Jet1-3). Mini-filament2 lies on N2 erupts and initiates the eruption of C-class flare and a blow-out jet (Jet4). The displayed HMI images are at respective initiation time of all four jets which are plotted in the top panel. The overplotted box on HMI magnetogram is used to estimate the magnetic fluxes at the base region of the recurring jets. The changing pattern of the positive and negative magnetic fluxes at the base region  of the recurring jets for their whole evolution period is shown in the bottom panels. The overplotted vertical lines on magnetic fluxes indicate the initiation times of all four jets.}
\label{Figure 3}
\end{figure*}
\begin{figure*}
\centerline{\includegraphics[width=13cm,angle=0]{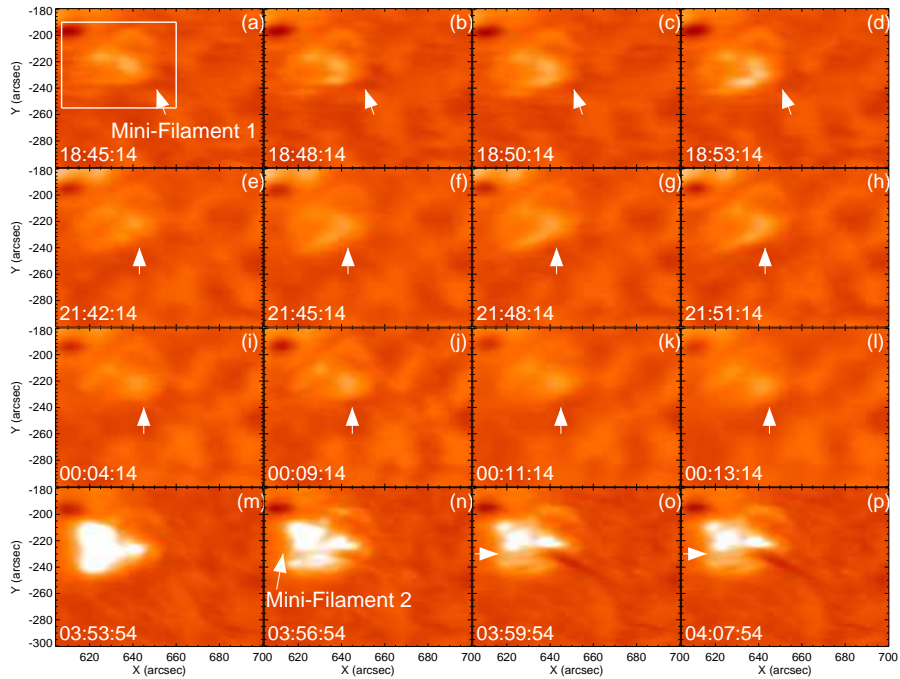}}
\caption{The mosaic shows the time-sequential images of H$\alpha$ for the eruption of the recurring jets (Jet1-4). In these images, we observe the presence of two mini-filaments at the base of the jets. Mini-filament1 is observed at the base of the homologous jets (Jet1-3). Mini-filament2 is observed at the base of Jet4. These two mini-filaments lie above two different neutral lines. The mini-filament at the base of first three jets partially erupt with the eruption of jets. In case of Jet4, the mini-filament2 fully erupts and makes a full blowout jet.}
\label{}
\end{figure*}

\begin{figure*}
\centerline{\includegraphics[width=13cm,angle=0]{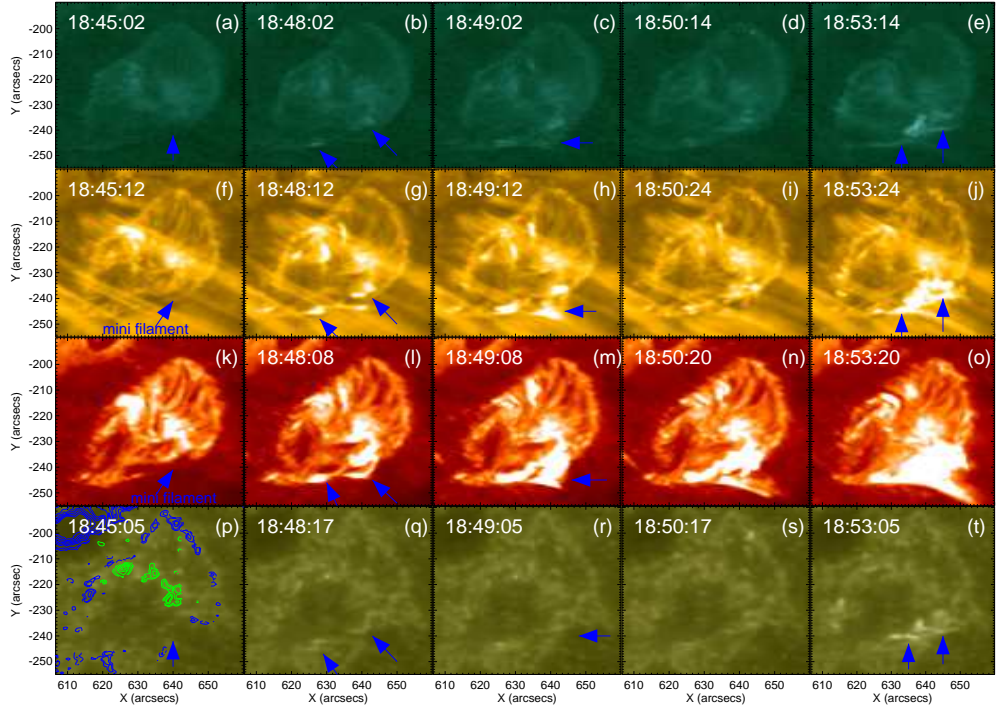}}
\caption{The mosaic shows the zoomed view of the base of Jet1 in multi-temperature filters of SDO/AIA (a-e: AIA 94 \AA~; f-j: AIA 171 \AA~, k-o: AIA 304 \AA~, p-t: AIA 1600 \AA~). At the initiation time of the jet, eruption of a mini-filament is present at the base of jet. This filament partially contributes in the eruption of Jet1. In image (p) of AIA 1600 \AA~, blue and green contour of HMI show the positive and negative magnetic polarities with the levels of $\pm{100}$, $\pm{200}$, $\pm{300}$, $\pm{500}$, $\pm{600}$, $\pm{700}$, $\pm{800}$, $\pm{900}$, $\pm{1000}$ G. The evolution of Jet1 in multi-temperature filters can be seen in animation Movie\_1.}

\label{}
\end{figure*}

\begin{figure*}
\centerline{\includegraphics[width=5cm,angle=90]{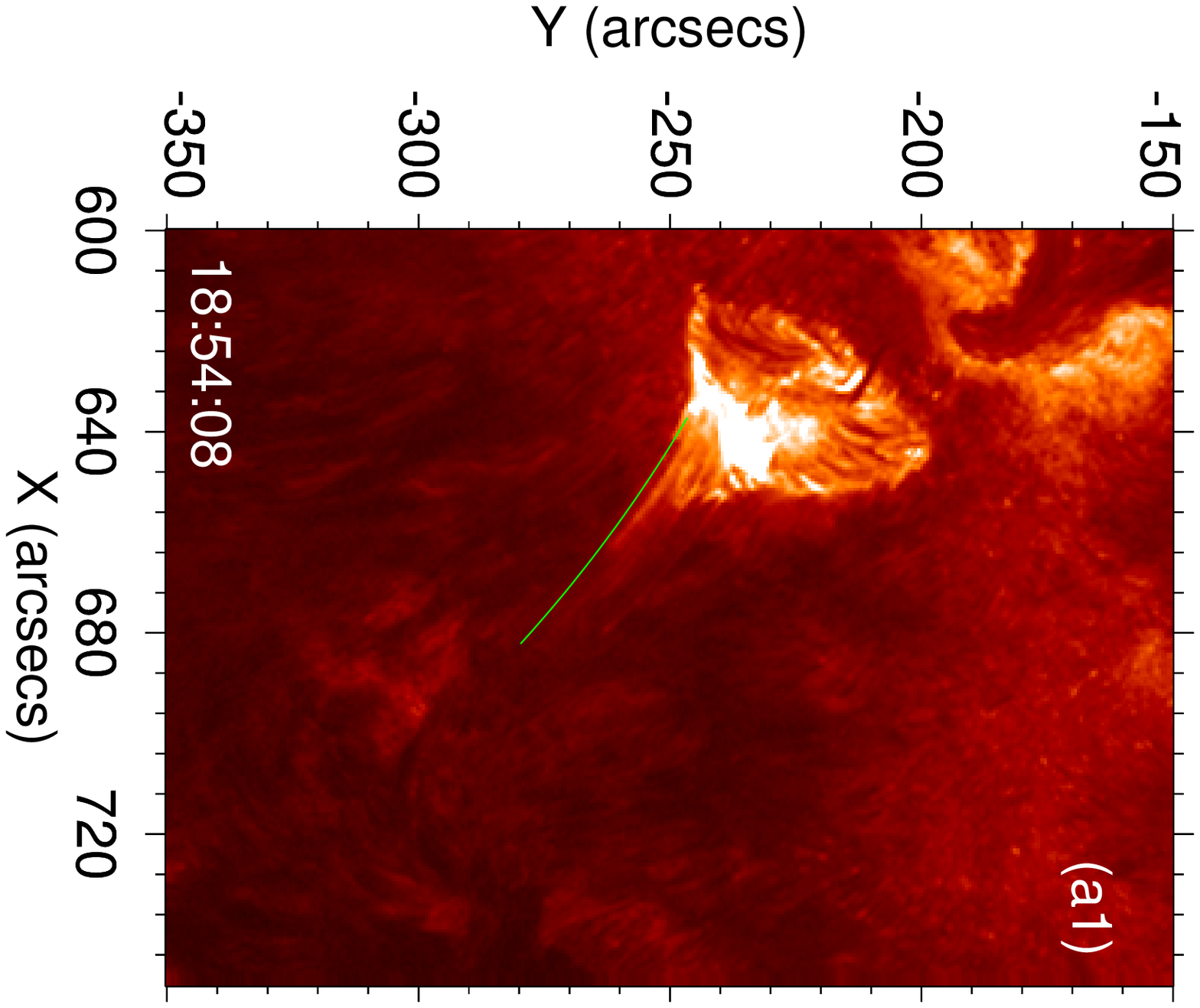}
\hspace*{-0.15\textwidth}
\includegraphics[width=5cm,angle=90]{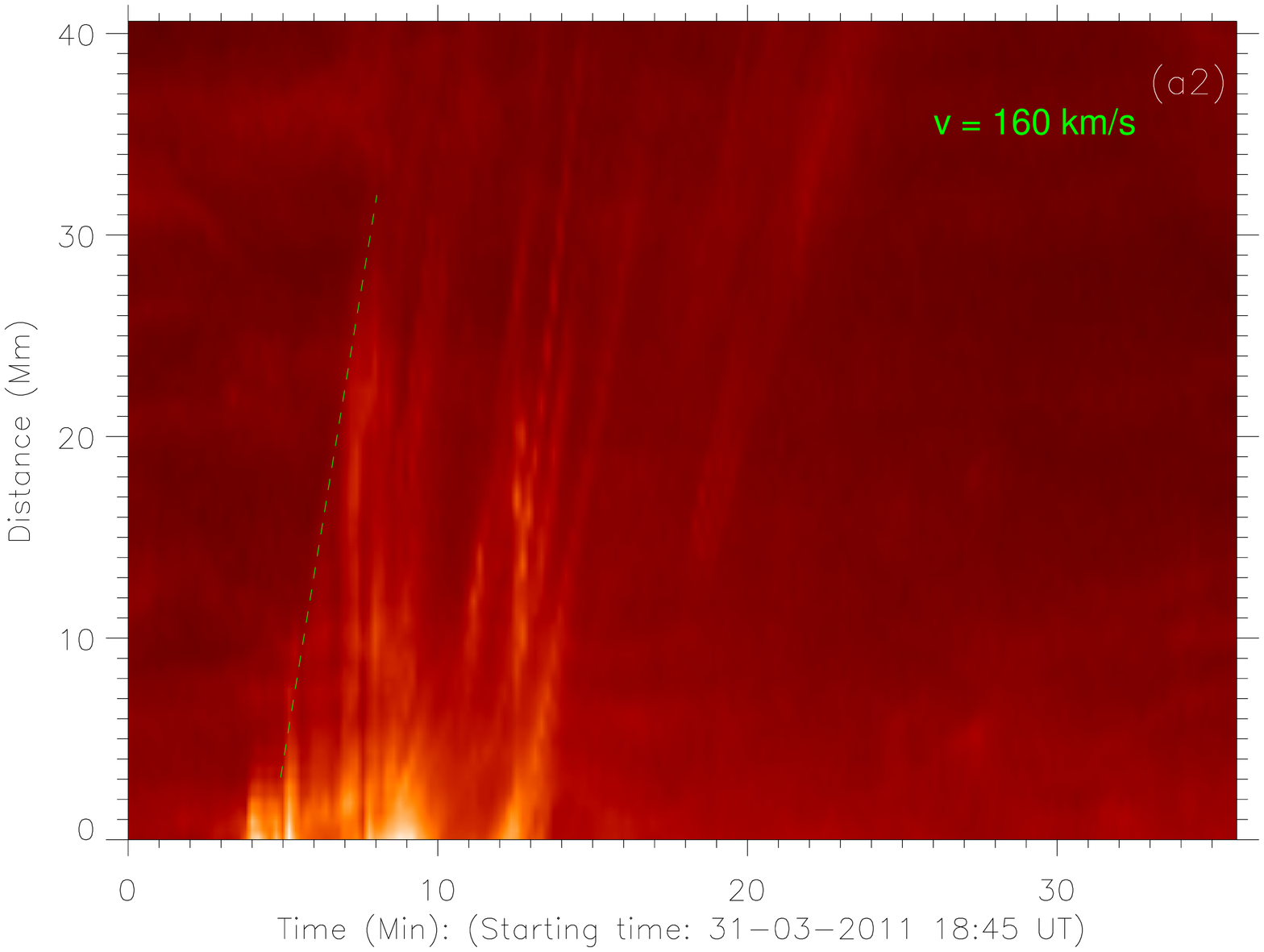}
}
\caption{The velocity estimation of Jet1 using the height-time analysis technique. The velocity is calculated along the overplotted path on image of AIA 304 \AA~ at 18:54:08 UT.}
\label{Figure 4}
\end{figure*}
\begin{figure*}
\centerline{\includegraphics[width=14cm,angle=0]{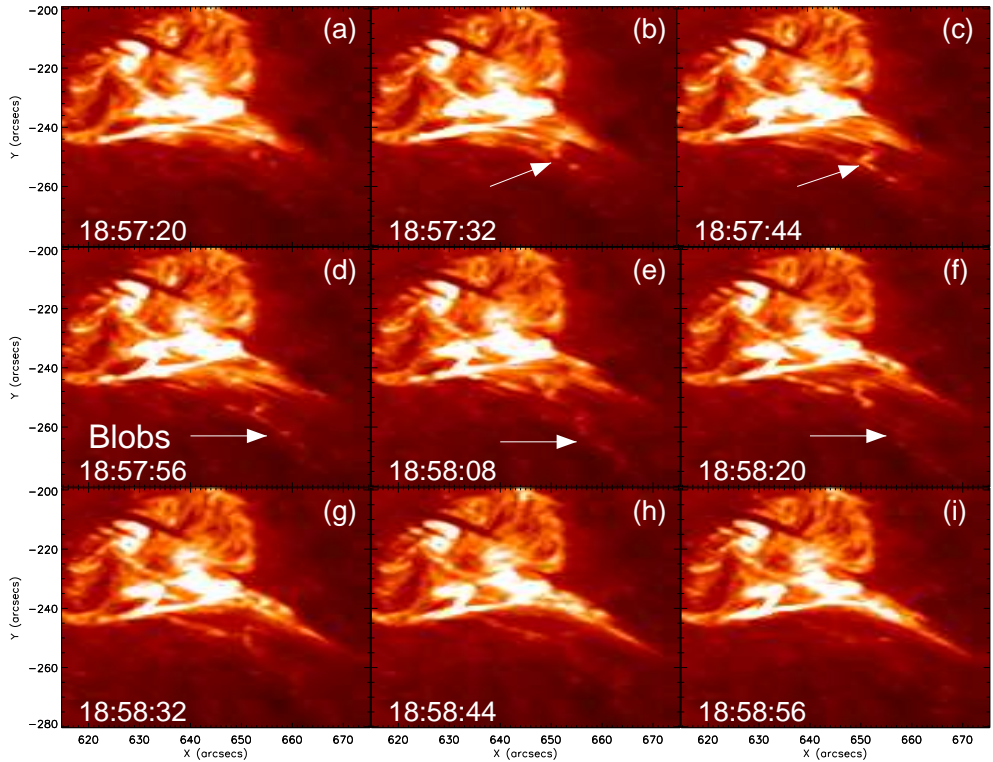}}
\caption{The mosaic in SDO/AIA 304 \AA~ shows the plasma blobs formation and their evolution in Jet1 eruption. The plasma blobs are formed in the spire of the jet and flow in the eruption. There are multiple blobs formation in the base of the jet due to magnetic reconnection which evolve at the time of the jet eruption. We indicate the multiple blobs by arrow on these images. These time-sequential images show the blobs movement along the jet spire. The base region is subjected to many dynamic fine loop threads coming out and interactive with each other. The reconnection between these threads leads the plasma blob eruption and the plasma flows along jet's spire.}
\label{Figure 5}
\end{figure*}
\begin{figure*}
\centerline{\includegraphics[width=14cm,angle=0]{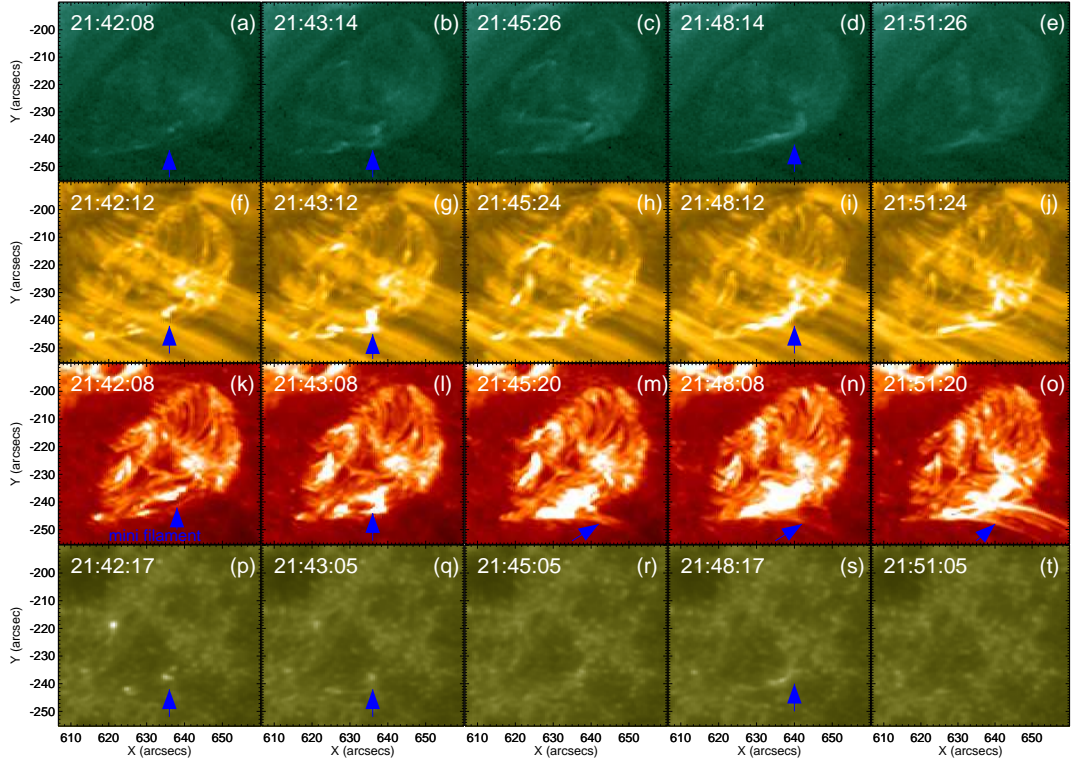}}

\caption{The behavior of Jet2 in multi-temperature filters of SDO/AIA (a-e: AIA 94 \AA~; f-j: AIA 171 \AA~, k-o: AIA 304 \AA~, p-t: AIA 1600 \AA~). The mini-filament at the base of the jet partially leaks with the eruption of Jet2. At initial time there is small bright loop structures formed at the base of the Jet2. The evolution of Jet2 in multi-temperature filters of SDO/AIA can be seen in Movie\_2.}
\label{Figure 8}
\end{figure*}
\begin{figure*}
\centerline{\includegraphics[width=5cm,angle=90]{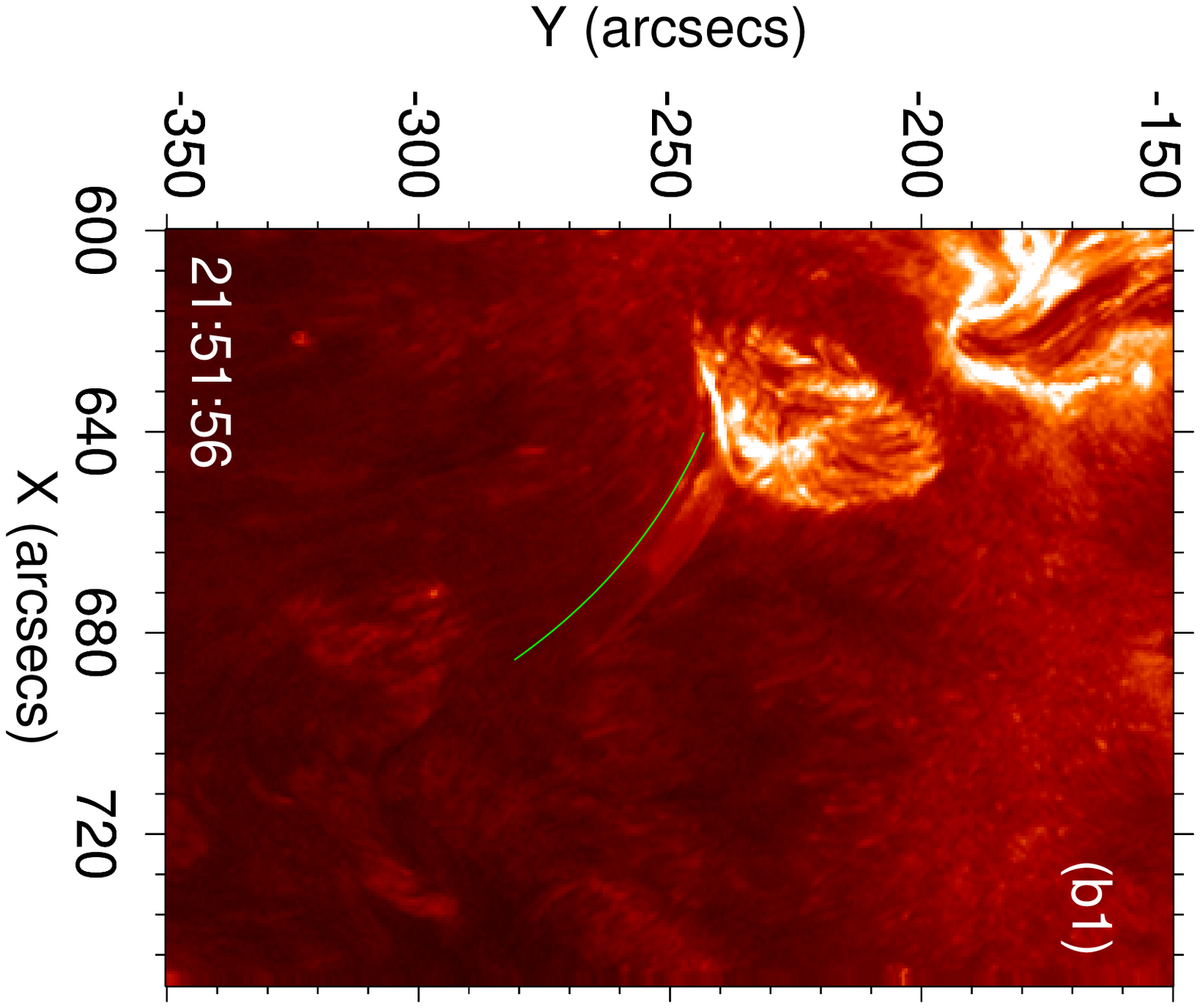}
\hspace*{-0.15\textwidth}
\includegraphics[width=5cm,angle=90]{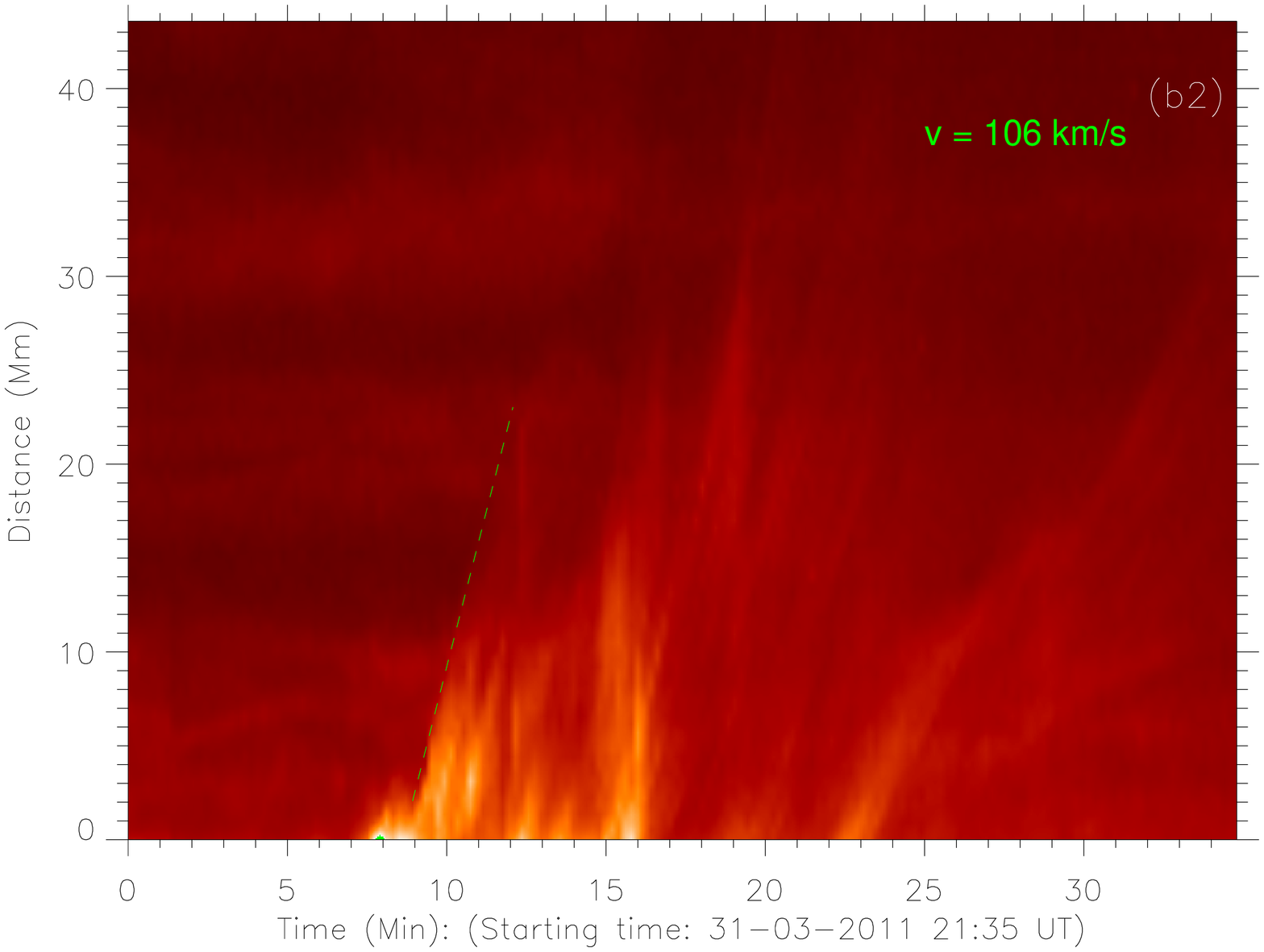}
}
\caption{The height-time measurement of Jet2 in SDO/AIA 304 \AA~ by using path as overplotted slit on image b1.}
\label{Figure 9}
\end{figure*}
\begin{figure*}
\centerline{\includegraphics[width=14cm,angle=0]{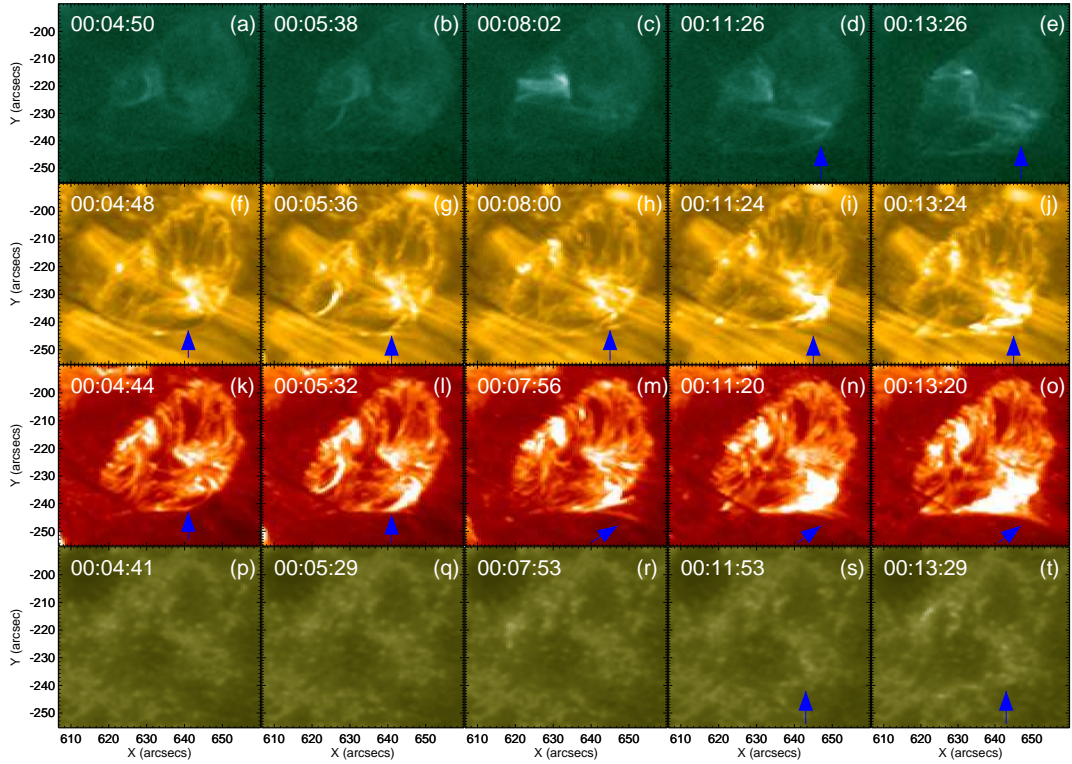}}
\caption{The mosaic of the zoomed view of eruption of Jet3 in multi-temperature filters of SDO/AIA (a-e: AIA 94 \AA~; f-j: AIA 171 \AA~, k-o: AIA 304 \AA~, p-t: AIA 1600 \AA~). This jet shows the very faint spire. In image (f) and image (k) blue arrow points towards the mini-filament at the base of jet, this mini-filament partially erupts with the eruption of Jet3. The evolution of Jet3 in multi-temperature filters of SDO/AIA can be seen in Movie\_3.}
\label{Figure 10}
\end{figure*}

\begin{figure*}
\centerline{\includegraphics[width=5cm,angle=90]{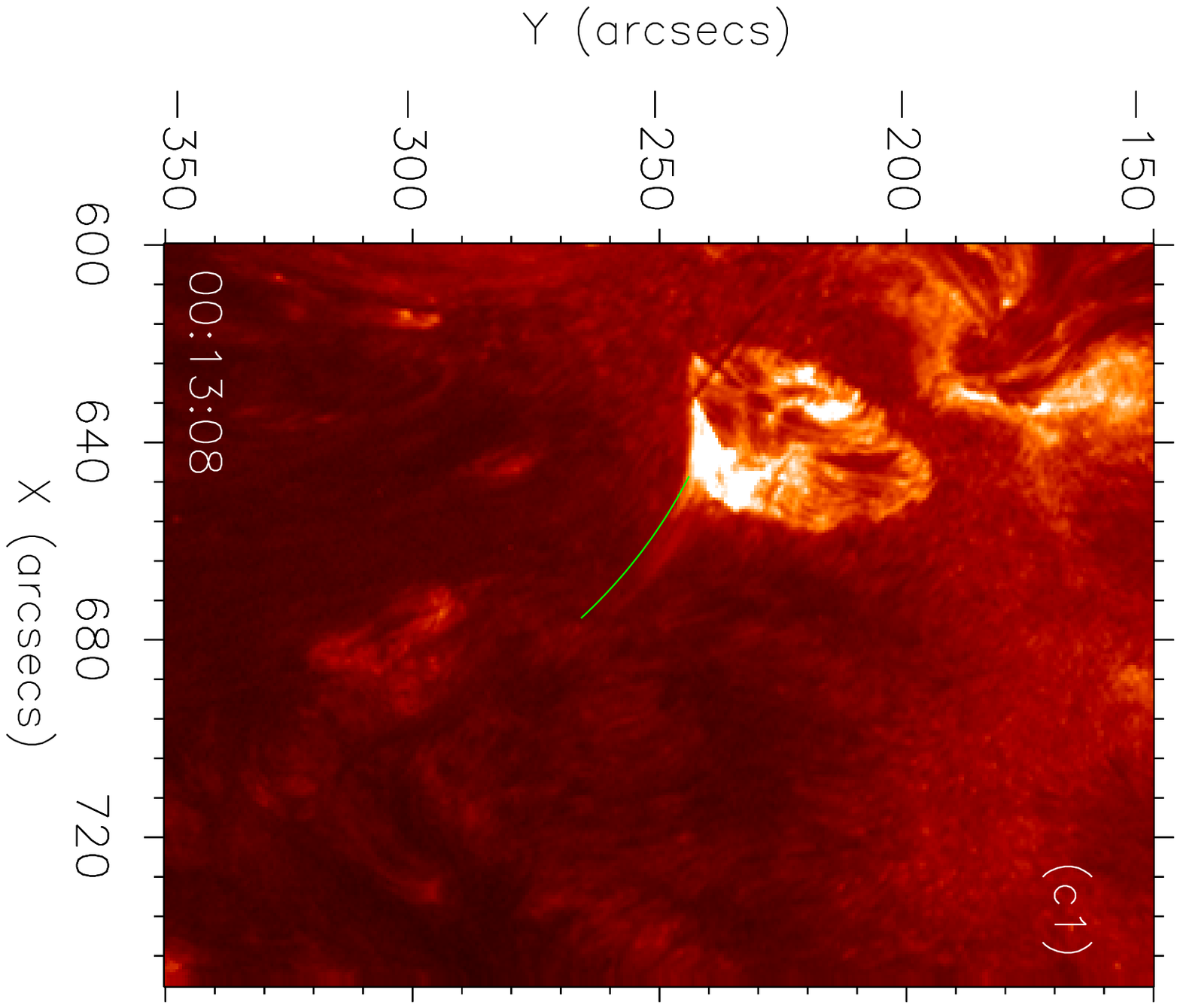}
\hspace*{-0.15\textwidth}
\includegraphics[width=5cm,angle=90]{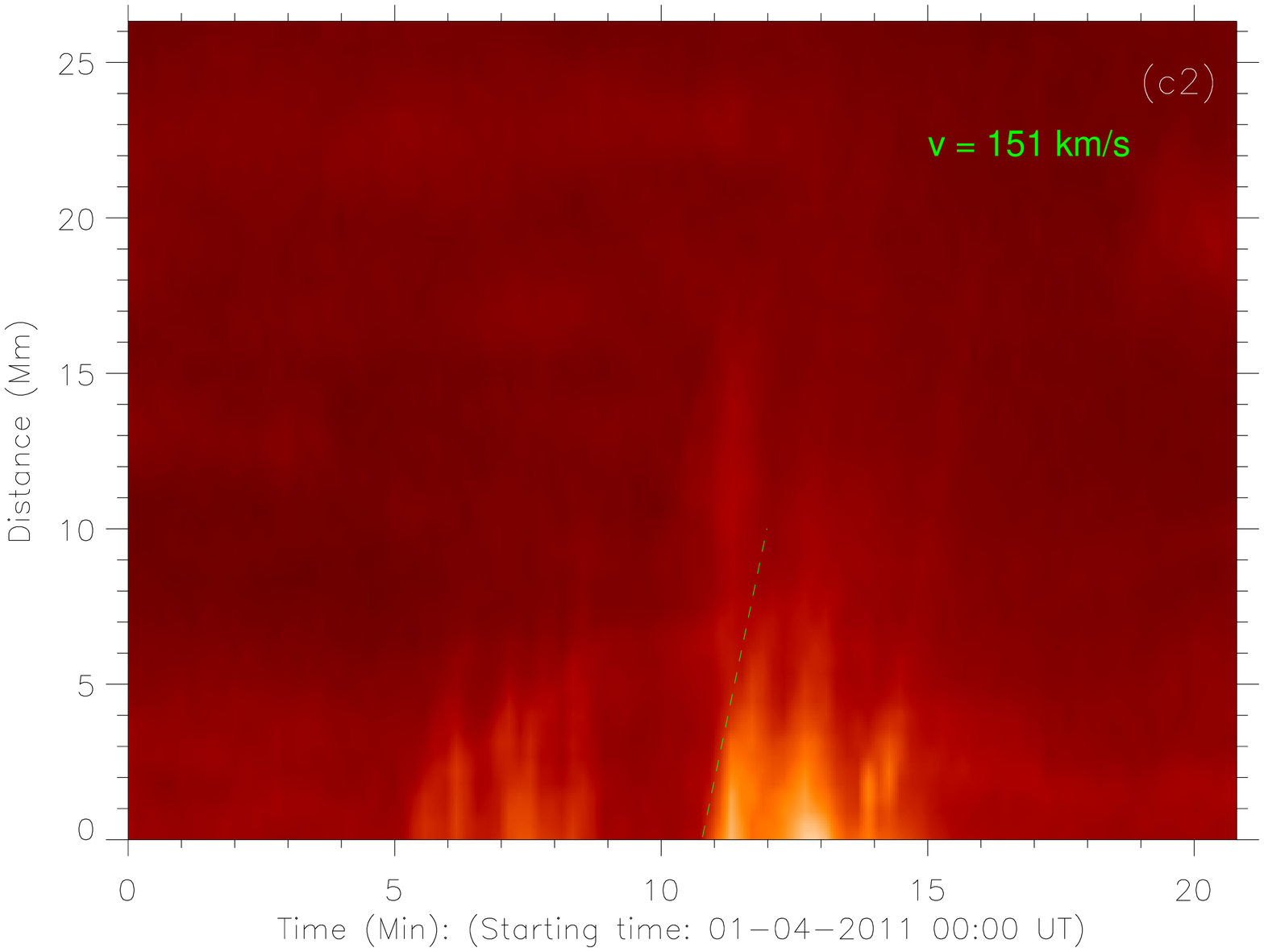}
}
\caption{The velocity of Jet3 is calculated along the overplotted slit on AIA 304 \AA~ image by using height-time measurement.}
\label{Figure 11}
\end{figure*}
\begin{figure*}
\centerline{\includegraphics[width=14cm,angle=0]{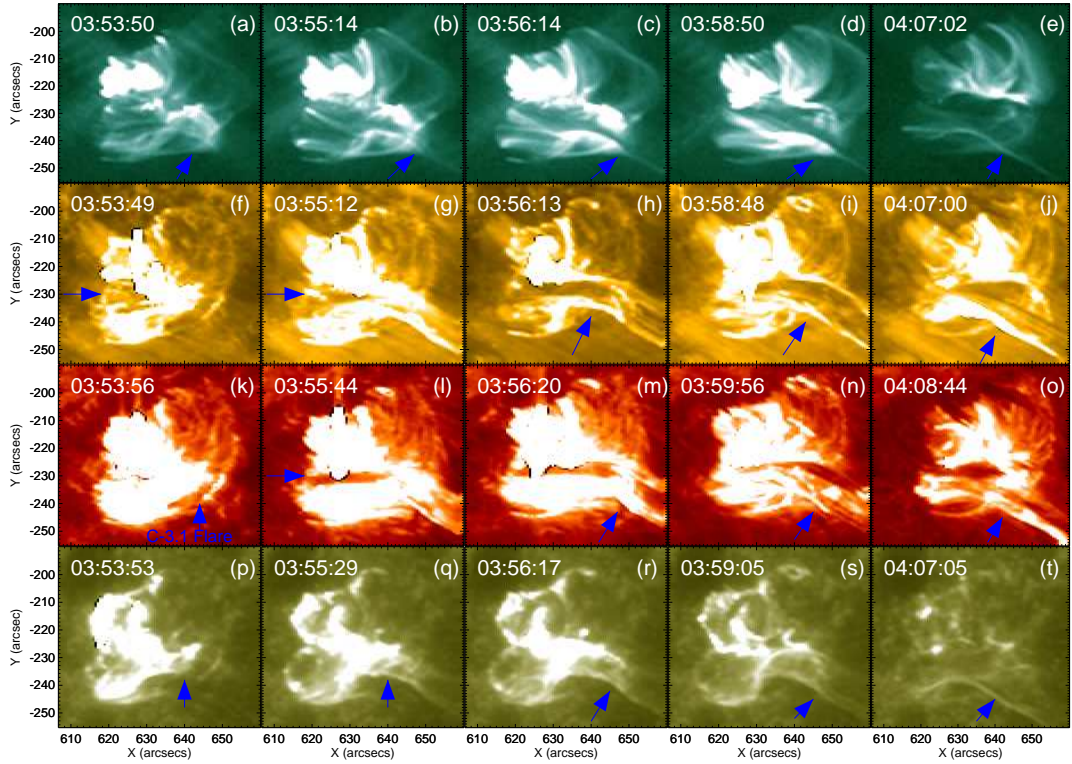}}

\caption{The evolution of large scale jet (Jet4) in the sequential images of multi-temperature filter of SDO/AIA (a-e: AIA 94 \AA~; f-j: AIA 171 \AA~, k-o: AIA 304 \AA~, p-t: AIA 1600 \AA~). Image (g) and image (l) points to the presence of the mini-filament at the base of the Jet4, this filament erupts and triggers the C-class flare and Jet4 eruption. Jet4 has spray like stucture. The long spire of the high-speed spray jet stays in its position for a long time and it directly converts to a CME. The eruption of Jet4 in multi-temperature filters of SDO/AIA can be seen in Movie\_4.}
\label{Figure 12}
\end{figure*}
\begin{figure*}
\centerline{\includegraphics[width=5cm,angle=90]{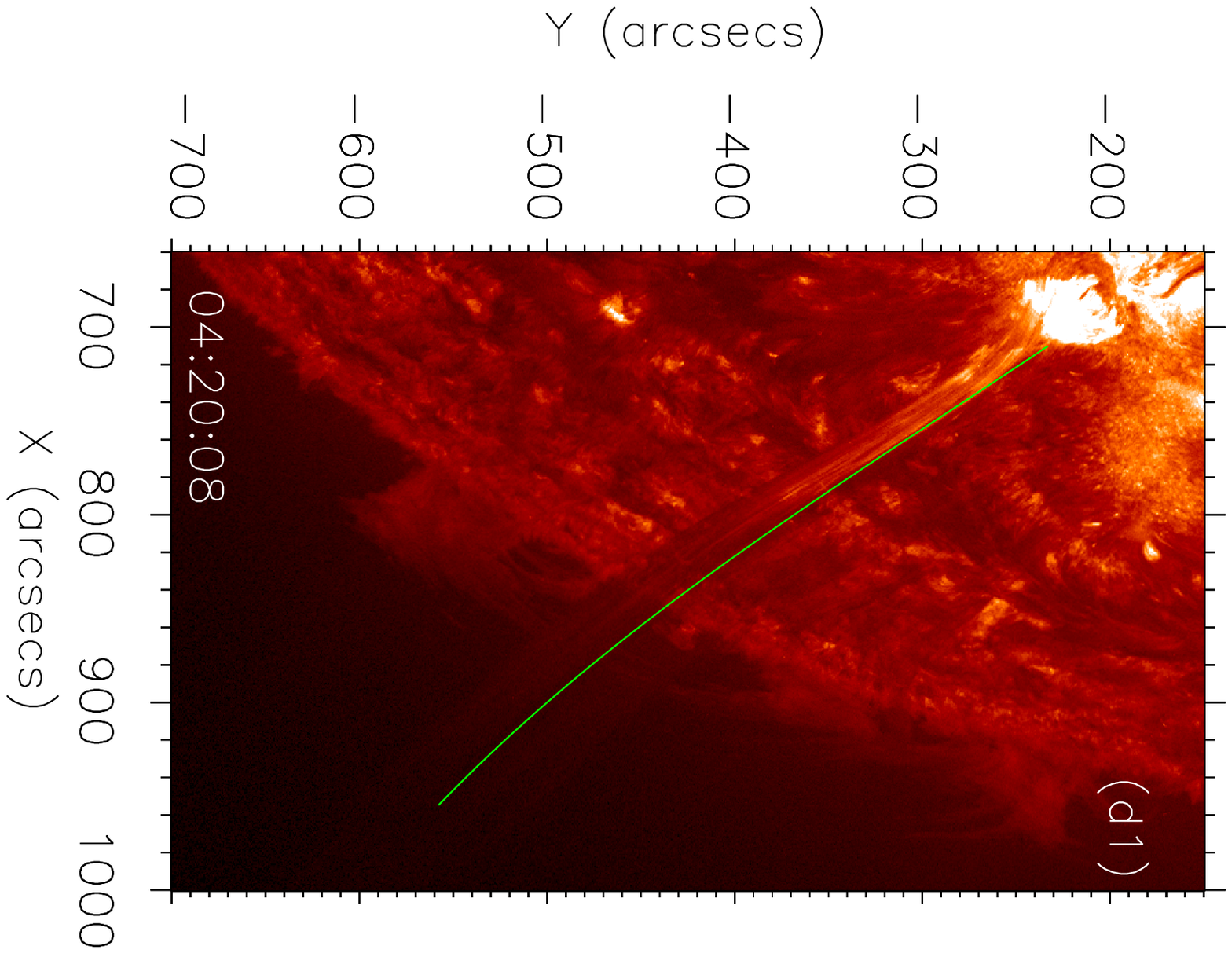}
\hspace*{-0.15\textwidth}
\includegraphics[width=5cm,angle=90]{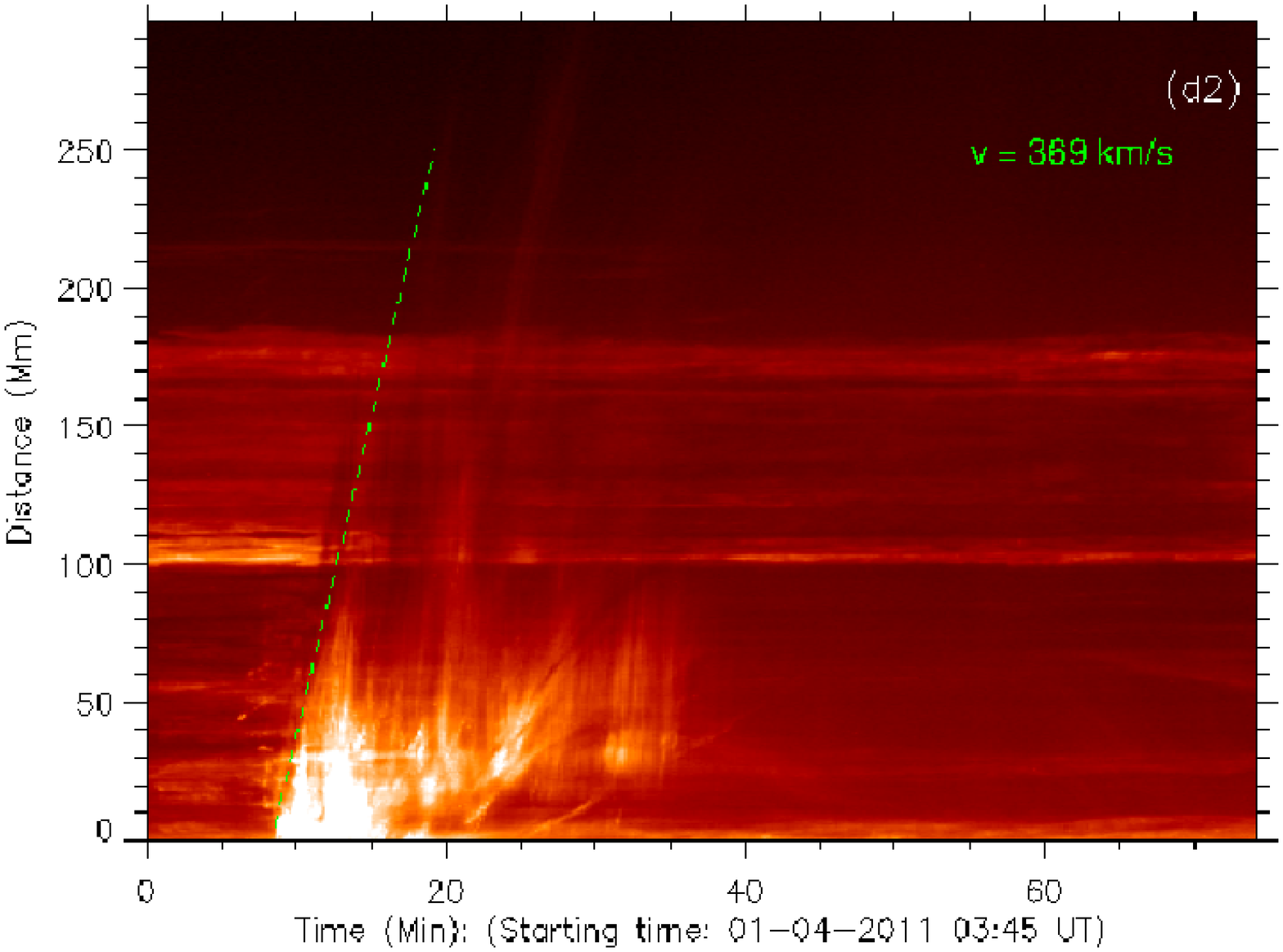}
}
\caption{Velocity estimation of Jet4 in AIA 304 \AA~ by using height-time measurement.}
\label{Figure 13}
\end{figure*}

\begin{figure*}
\centerline{\includegraphics[width=9.8cm,angle=90]{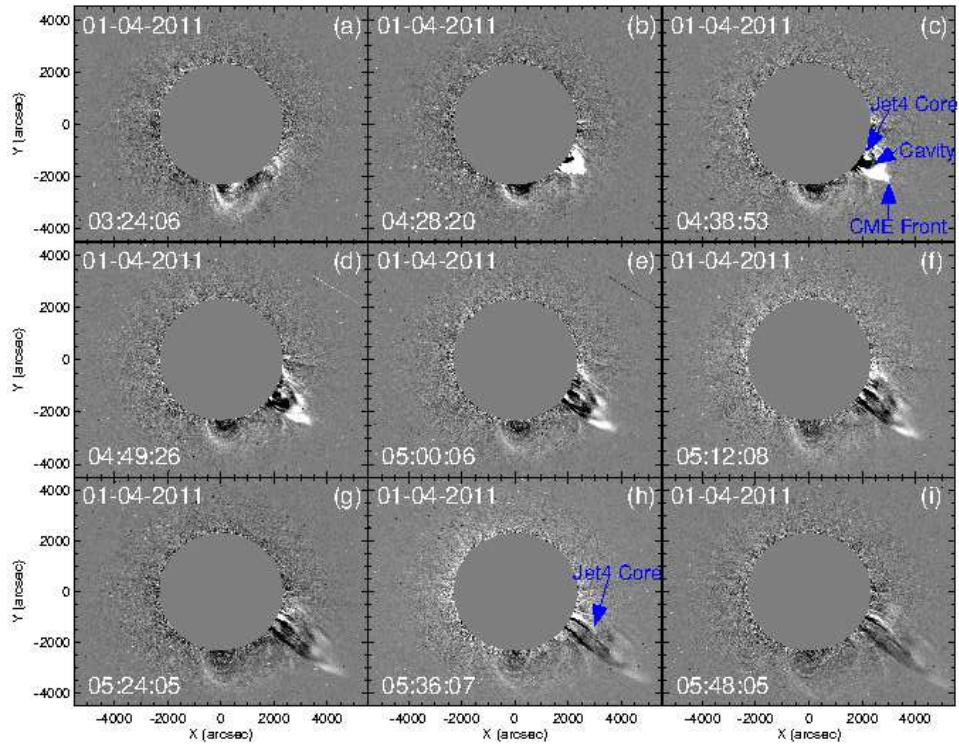}}
\caption{The mosaic of time-sequential images of LASCO C2 coronagraph shows the eruption of Jet4 associated CME. The evolution of Jet4 associated CME can be seen in Movie\_5.}
\label{Figure 14}
\end{figure*}
\begin{figure*}
\centerline{\includegraphics[width=10.0cm,angle=0]{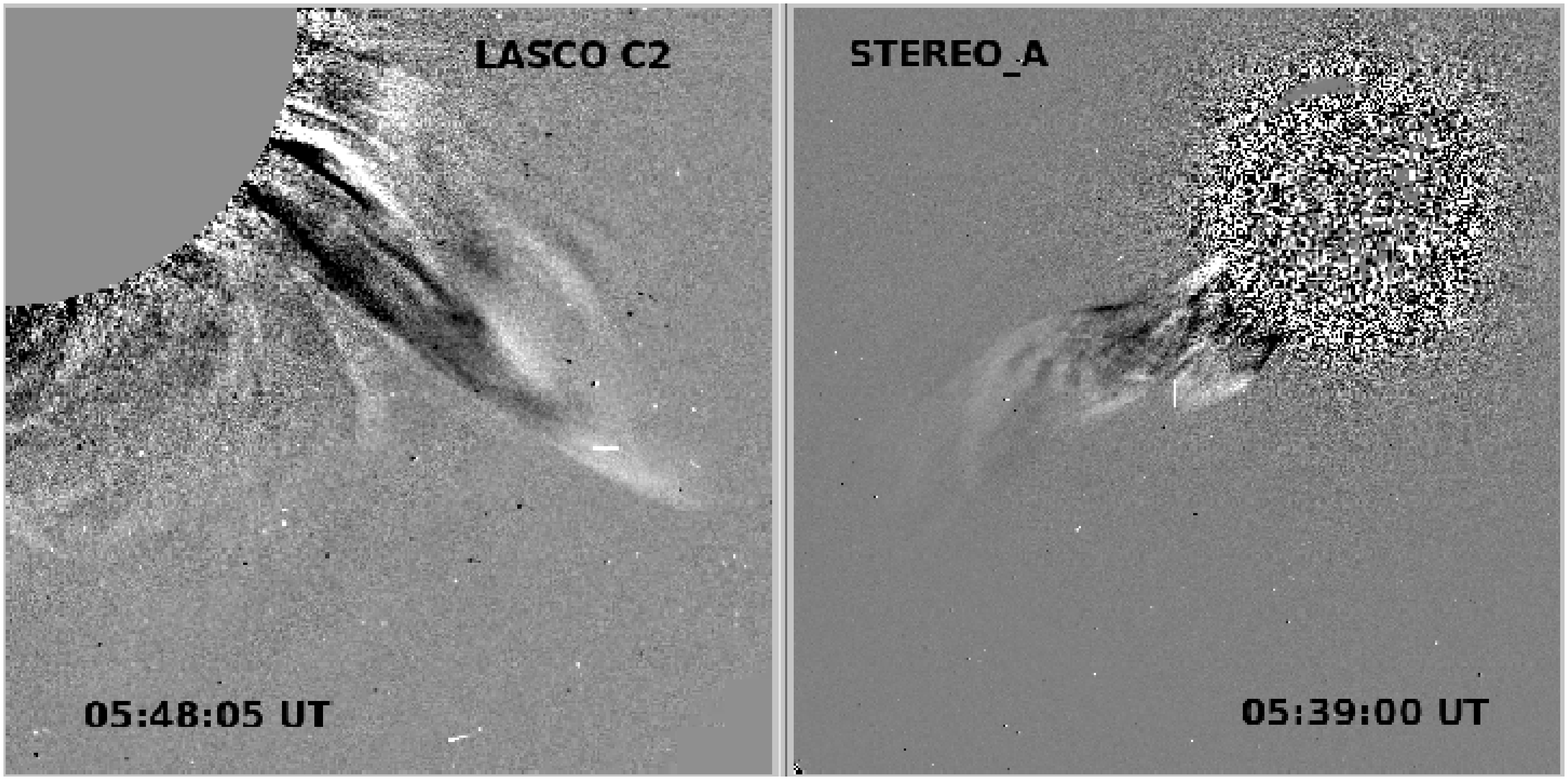}}
\vspace*{0.04\textwidth}
\centerline{\includegraphics[width=7.6cm,angle=90]{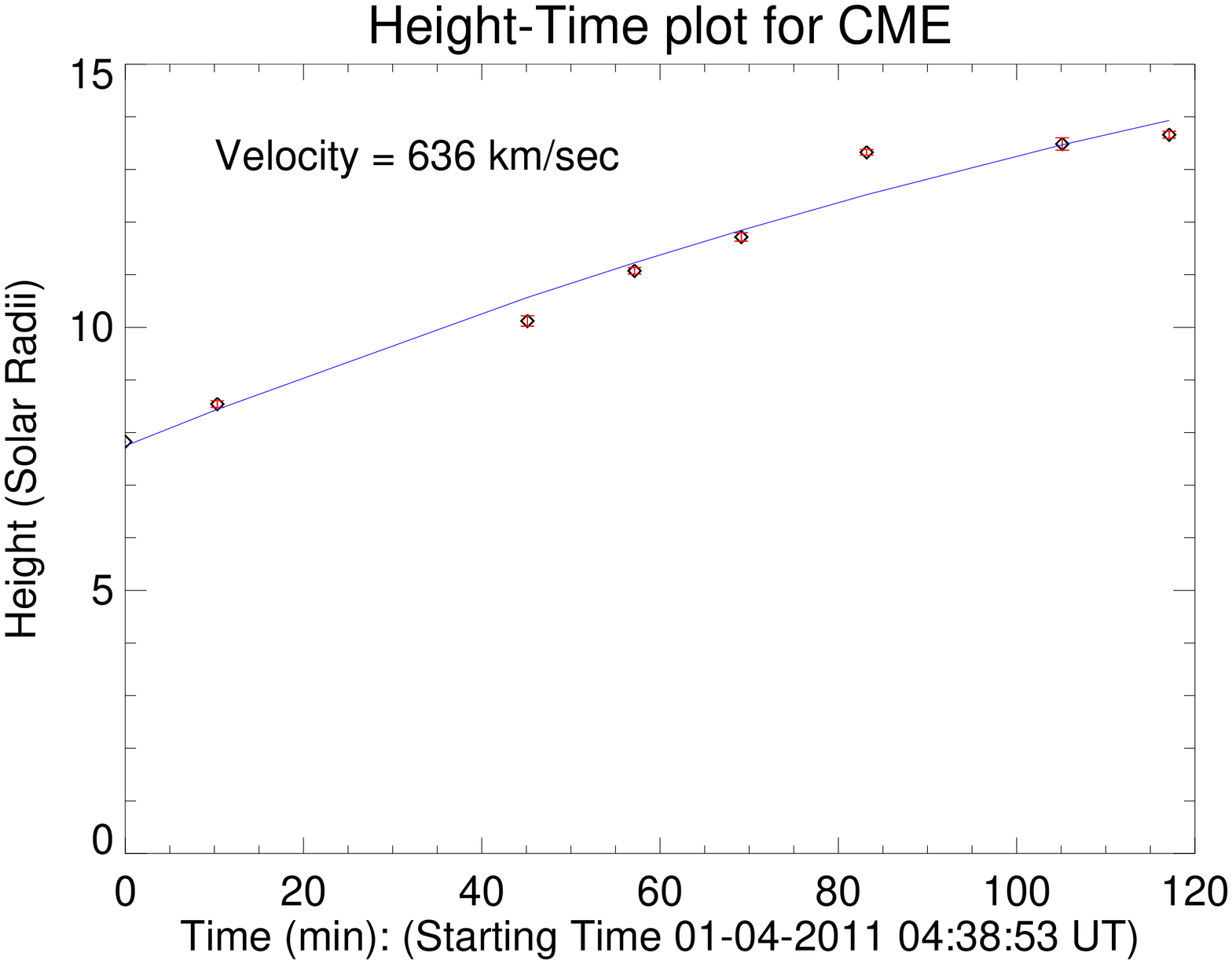}
}
\caption{The tracking the tip of the CME in STEREO\_A COR2 and LASCO C2 with the help of triangulation method. The calculated velocity of CME is $636$ $km$ $s^{-1}$ by using the height-time measurement technique.}
\end{figure*}
\label{Figure 15}
\begin{acks}
R.S. acknowledges the Department of physics, Indian Institute of Technology (BHU) for providing her Senior Research Fellowship (SRF) and computational facilities. Authors acknowledge the space-based observations of SDO/AIA, SDO/HMI, SoHO/LASCO and STEREO for this work. AKS acknowledges the support from UKIERI grant to conduct his scientific research.
\end{acks}
\newline
\textbf{Declaration of Potential Conflicts of Interest} Authors declare that they have no conflicts of interest.

\end{article} 
\end{document}